\documentclass[a4paper,12pt]{article}
\pdfoutput=1 

\usepackage{jheppub} 

\usepackage[T1]{fontenc} 
\usepackage{physics}
\NewDocumentCommand{\tens}{t_}
{%
	\IfBooleanTF{#1}
	{\tensop}
	{\otimes}%
}
\NewDocumentCommand{\tensop}{m}
{%
	\mathbin{\mathop{\otimes}\displaylimits_{#1}}%
}
\usepackage{graphicx}
\usepackage{wrapfig}
\usepackage{amsmath}
\usepackage{amsfonts}
\usepackage{amssymb}
\usepackage{quotes}
\usepackage{subcaption}
\usepackage{calligra}
\newcommand{\rc}{\textit{\huge\calligra r}}
\usepackage{changepage} 

\usepackage{afterpage}

\usepackage{placeins}
\newcommand{\z}{\Bar{z}}
\newcommand{\paren}[1]{\left(#1\right)}
\newcommand{\m}{\Tilde{m}}
\newcommand{\C}{(l^2+\Bar{z}^2)}
\newcommand{\ci}{(l^2+z_i^2)}
\newcommand{\T}[1]{\tilde{#1}}

\title{\textbf{\boldmath Holographic entanglement entropy and complexity for the cosmological braneworld model}}

\author{Souvik Paul,}
\author[1]{Gopinath Guin,\note{Corresponding author.}}
\author{Sunandan Gangopadhyay}

\affiliation[a]{\textit{Department of Astrophysics and High Energy Physics\\}
	\textit{ S.N.~Bose National Centre for Basic Sciences,\\}
	\textit{JD Block, Sector-III, Salt Lake, Kolkata 700106, India}}

\emailAdd{souvik.paul@bose.res.in}
\emailAdd{gopinath.guin@bose.res.in}
\emailAdd{sunandan.gangopadhyay@bose.res.in}

\abstract{\noindent In a recent study \cite{Park:2020jio}, the time-dependent entanglement entropy of the universe undergoing expansion according to various power laws has been analyzed within the framework of the braneworld model. The results of the entanglement entropy in that paper take into account only the effects of a radiation and a matter-dominated universe. In this work, we have computed the time-dependent entanglement entropy and complexity of the FLRW universe in the presence of different matter sources (radiation, matter and some exotic matter). In contrast to the approach in  \cite{Park:2020jio}, all the calculations in this paper have been carried out in a perturbative manner in the framework of braneworld model of cosmology. 
	According to this model, our universe is situated on a brane and different matter sources appear on the brane due to the back reaction of different $p$-brane gas configurations in the bulk spacetime. By considering the bulk spacetime as a black brane geometry, we have considered different blackening factors corresponding to radiation, matter, and exotic matter and calculated entanglement entropy and complexity holographically using the methods presented in the literature. In the braneworld model, the universe's expansion is described by the brane's time-dependent radial position. This position is determined using the second Israel junction condition for various matter sources. The time evolution of entanglement entropy and complexity is then obtained by substituting this brane position.  We have also shown the dependence of entanglement entropy and complexity on the cosmological time for all the different matter-dominated universes in the early and late time eras. Even though all the calculations of holographic entanglement entropy and complexity are done in a perturbative way, the early and late time behaviour of holographic entanglement entropy matches with \cite{Park:2020jio}.}
\begin{document}
	\maketitle
	\flushbottom
	\section{Introduction}
	Over the past few years, the AdS/CFT duality \cite{Maldacena:1997re,Gubser:1998bc,Witten:1998qj,Aharony:1999ti,Nastase:2007kj,Natsuume:2014sfa} has gained attention by the scientific community for its ability to study strongly coupled gauge theories with the help of a weakly coupled gravity theory having one extra dimension. This idea initially came up while studying the large $N$ limit of $SU(N)$ gauge theories. In large $N$ limit $SU(N)$ theories behave like string theories with one extra dimension and having a coupling constant of $1/N$ \cite{tHooft:1973alw}. It was first shown by Maldacena that the large $N$ limit of a $\mathcal{N}=4$ super Yang-mills theory is dual to a type-IIB superstring theory \cite{Maldacena:1997re}. From their this duality has been used in several branches of physics, like 
	black hole physics \cite{Emparan:2002px,Tanaka:2002rb,Gregory:2004vt}, quantum gravity \cite{Bak:2006nh,Rovelli:1997na,Silva:2023ieb,Engelhardt:2015gla}, QCD \cite{Erlich:2005qh,Karch:2006pv,Kruczenski:2004me,Andreev:2006ct,Csaki:2008dt}, condensed matter \cite{Hartnoll:2008kx,Hartnoll:2008vx,Herzog:2009xv,Li:2011xja,Gangopadhyay:2012am,Horowitz:2010gk} etc. Later this duality is used to study different information theoretic quantities of different CFTs holographically. This idea to calculate different quantum information theoretic quantities holographically was first proposed in \cite{Ryu:2006bv,Ryu:2006ef}. In this paper we will study few quantum information theoretic properties of the Friedmann–Lemaitre–Robertson–Walker (FLRW) \cite{Friedman:1922kd,Friedmann:1924bb,Lemaitre:1931zza,Lemaitre:1933gd,Robertson:1935jpx,Robertson:1935zz,Walker:1937qxv} universe in presence of different matter sources holographically. Before doing so we will briefly discuss about entanglement entropy and complexity from the context of quantum information theory.\\
	In the context of quantum information theory the concept of entanglement entropy is a very important quantity. The entanglement entropy measures the degrees of quantum entanglement between two subsystems, which forms a two-part quantum system. Now for a pure bipartite quantum state of a composite system the entanglement entropy is nothing but the von Neumann entropy of the reduced density matrix of any of the systems. In order to calculate the von Neumann entropy in the context of quantum information theory, we need to follow the following procedure. Let us choose a pure bipartite quantum system with Hilbert space of the form $\mathcal{H}=\mathcal{H}_{A}\tens\mathcal{H}_{B}$, where $A$ and $B$ is the full system. The density matrix of the total system is given by $\rho_{AB}=\ket{\psi}\bra{\psi}$, where $\ket{\psi}\in \mathcal{H}$. The entanglement entropy of a given system $A$ is obtained by calculating the von Neumann entropy of the reduced density matrix, which is obtained by integrating over the degrees of freedom outside $A$. Thus, the von-Neumann entropy of a subsystem $A$ is given by\cite{neumann2013mathematische,nielsen2010quantum}
	\begin{equation}
		S_{A}=-\Tr(\rho_{A}\log \rho_{A})
	\end{equation}
	where $\rho_{A}$ is the reduced density matrix of the subsystem $A$. This reduced density matrix is obtained by tracing over the degrees of freedom of subsystem $B$. Symbolically one can denote $\rho_{A}=\Tr_{B}\rho_{AB}$.However for mixed states the von Neumann entropy is not a good measurement of entanglement due to the involvement of irrelevant classical correlations. Different correlation measures (like mutual information \cite{nielsen2010quantum,Wolf:2007tdq}, entanglement of purification \cite{Terhal:2002riz,Takayanagi:2017knl} and entanglement negativity \cite{Vidal:2002zz,Plenio:2005cwa,Zyczkowski:1998yd}) have been proposed in the previous literature. As well as other mixed-state entanglement measures, quantum complexity \cite{Jefferson:2017sdb,Chapman:2017rqy,Hackl:2018ptj,Bhattacharyya:2018bbv,Khan:2018rzm} has gained a notable focus. For pure states, complexity is defined as the minimum number of gates that are chosen from a pre-established set of gates which takes us from a reference state ($\ket{\psi}_{R}$) to a target state ($\ket{\psi}_{T}$). If $\ket{\psi}_{T}$ is represented as $\ket{\psi}_{T}=\hat{U}\ket{\psi}_{R}$, where $\hat{U}$ is the unitary operator that transforms reference state to target state. This operator $\hat{U}$ is constructed from a set of elementary gates, thus one can write
	\begin{equation}
		\hat{U}=g_{1}g_{2}g_{3}\dots g_{n-1}g_{n}~.
	\end{equation}
	where $g_{i}$ denotes the elementary quantum gate. If the circuit complexity of the target state is denoted by $C(\ket{\psi}_{T})$, then it is defined as the smallest number of gates required to create the unitary operation. This is one of the ways to compute quantum complexity, although there are other techniques to calculate complexity as mentioned in \cite{Nielsen:2005mkt,Nielsen:2006cea,Ali:2018fcz,Susskind:2018pmk}. For strongly interacting systems computation of quantum complexity is a bit difficult, although holographic techniques make the calculation much easier. In the upcoming sections of this paper we have discussed several holographic techniques to compute complexity.\\
	In this paper, we have computed the entanglement entropy and complexity of the FLRW universe using holographic duality. Before the FLRW model came into the picture, Einstein proposed a static model \cite{einstein1917kosmologische,ORaifeartaigh:2017uct} considering the existence of homogeneity and isotropy in the universe. In this model, the universe has no beginning and expansion. At that time de Sitter also proposed a model \cite{deSitter:1917zz} of the universe that deals with static cosmological solutions with the astrophysical bodies and the system considered to be test particles. In this model, astrophysical bodies are capable of affecting the geometry of the spacetime. However, these models proposed by Einstein and de Sitter have some inconsistencies because Einstein's model does not consider the red shift in the universe, on the other hand, de Sitter's model does not have any kind of matter. It was the FLRW model \cite{Friedman:1922kd,Friedmann:1924bb,Lemaitre:1931zza,Lemaitre:1933gd,Robertson:1935jpx,Robertson:1935zz,Walker:1937qxv} which considered a non-static solution of the universe. In this model, the red shift of the light sources of distant parts of the universe can also be explained. Later in \cite{eddington1930instability}, Eddington showed the instability of Einstein's static model of the universe. Eddington's argument was also supported by Hubble's observations \cite{Hubble:1929ig}. Thus, it was confirmed that our universe is homogeneous, isotropic but non-static in nature. On the other hand in cosmology, the dynamics of the gauge fields occur in an expanding background. After almost ten microseconds of the Big Bang when the temperature of the universe crossed the critical temperature of QCD, some strongly coupled process took place. In this case, the AdS/CFT framework to study the universe can be useful. Another theoretical tool in the context of cosmology is the ds/CFT correspondence \cite{Witten:2001kn,Strominger:2001pn,Maldacena:2002vr}. As our FLRW universe is expanding, to study the universes' entanglement properties, we need to use the Hubeny-Rangamani-Takayanagi (HRT) \cite{Hubeny:2007xt} formula instead of the RT formula. This HRT formula is sometimes called the covariant formulation. Although there is a problem in using the HRT formula, except few simpler cases it is difficult to exactly calculate the time-dependent entanglement entropy. Although it is known that in the UV limit, the RT formula gives the leading order contribution to the HRT formula in a given time \cite{Koh:2018rsw}. There are still some questions about the higher-order corrections to the late time of the entanglement entropy and how one can calculate the entanglement entropy of the universe expanding by a power law. The de Sitter boundary model of cosmology explains the properties of the eternally inflating universe \cite{Lowe:2004zs,Lowe:2010np}. But this model does not deal with the universe expanding by power law. In order to calculate the entanglement entropy of the expanding FLRW universe, researchers have used techniques in which through an appropriate coordinate transformation the AdS space is sliced in a way that the boundary spacetime becomes FLRW like \cite{Giataganas:2021cwg,Giantsos:2022qdd}. An alternative framework for exploring the universe through a power law expansion is the braneworld (or Randall-Sundrum) model of cosmology \cite{Randall:1999vf,Chamblin:1999by,Brax:2004xh,Chamblin:1999ya,Lee:2007ka,Randall:1999ee,Brax:2003fv,Flanagan:1999dc,Coley:2001ab}. According to this model our universe is situated on a brane which is embedded in a higher dimensional spacetime. Initially, this braneworld model was used to fix the hierarchy issue of two theories having different energy scales. Later this theory was improved to study inflationary cosmology with a graceful exist\cite{Randall:1999vf,Park:2000ga,Kraus:1999it}.\\
	Recently this braneworld model has been used to calculate the entanglement entropy of the universe using the holographic principle \cite{Park:2020jio}. In this paper, the RT formula is used because, in the braneworld model, the expansion of the universe is equivalent to the motion of the brane in the bulk direction. 
	The time-dependent radial position of the brane is determined by the second Israel junction condition. In this paper, the author has shown the effects of different matter fields on the holographic entanglement entropy using the RT formula. Then, the time dependence is incorporated in the result of entanglement entropy by substituting the time-dependent radial position of the brane in the results of HEE. In the braneworld model, different matter sources appear due to the back reaction of p-brane gas in the bulk \cite{Park:2021wep}. In that paper, the time-dependent entanglement entropy is calculated for radiation and matter-dominated eras. Although the effect of some kind of exotic matter was ignored in the previous literature. In this paper, we have calculated the time-dependent entanglement entropy and volume complexity of the FLRW universe expanding by a power law and containing three different kinds of matter sources (radiation, matter and some kind of exotic matter). It should be mentioned that our analysis varies from the method of \cite{Park:2020jio} in a way that we have done all of our calculations in a perturbative manner. In the braneworld model, these matter sources appear due to the back reaction of different kinds of $p$-brane gas geometries in the bulk. For example, 0-branes in the bulk correspond to radiation, 1-brane corresponds to matter, 2-brane corresponds to exotic matter (also referred to as one-dimensional cosmic strings \cite{Hindmarsh:1994re,Wu:1998mr,Sakellariadou:2009ev,Sakellariadou:2006qs,Shellard:1987bv,Vilenkin:1984ib}) and 3-brane corresponds to domain walls \cite{Vilenkin:1984ib,Bergshoeff:2003sy,Bremer:1998zp,Cvetic:2000pn,Vilenkin:1982ks,Kachru:2000hf,Blanco-Pillado:2022rad,Cvetic:1996vr,Boonstra:1998mp} in the braneworld model of cosmology. We have used the RT formula and "Complexity=Volume" conjecture to calculate the holographic entanglement entropy (HEE) and holographic subregion complexity (HSC) for a small circular subsystem on the brane for an eternally inflating universe and a universe expanding by power laws in the presence of radiation, matter and exotic matter. To calculate the time dependence of HEE and HSC, we have substituted the time-dependent brane positions for different matter sources in the universe. The time-dependent radial position is calculated using the second Israel junction condition. We have calculated the leading order time dependence of HEE and HSC for matter, radiation and exotic matter dominated the universe. We have also shown the early and late time behaviours of these quantities for different matter sources. We have also graphically represented all our results of HEE and HSC for different matter-dominated universes. Although we have done a complete perturbative calculation of HEE for different matter sources, the leading order early and late time dependence of HEE for radiation and matter dominated universe matches with the results of HEE in \cite{Park:2020jio}.\\
	This paper is organised as follows. In section \eqref{The Braneworld model in cosmology}, we have given a brief review of the braneworld model of cosmology. In section \eqref{Cosmological p-brane gas geometry model} we have shown how different $p$-brane gas configurations give rise to different matter sources in the universe. The time-dependent radial position of the brane is calculated for different matter-dominated eras of the universe using the Israel junction condition. In Section \eqref{Entanglement entropy of the universe}, we have computed the early and late time behaviour of the time-dependent entanglement entropy of the universe. Then we have calculated the same for the universe expanding by power law in radiation, matter and exotic matter-dominated eras. In Section \eqref{Complexity of the universe}, we have calculated the time-dependent volume complexity of the universe using the "Complexity=Volume" conjecture. The calculations are done for an eternally inflating universe and a universe with different matter-dominated universes in early and late times. We finally conclude in section \eqref{sec 7}. We have also provided a short introduction to the FLRW universe and discussed different matter sources present in the universe expanding by a power law in appendix \eqref{appendix}.
	\section{The Braneworld model in cosmology}\label{The Braneworld model in cosmology}
	In this section, we will discuss the braneworld model in cosmology. In this model, it is assumed that our four-dimensional universe is situated on a brane, which is embedded in a bulk with one extra dimension. Therefore, the radial motion of the brane with time is assumed to correspond the expansion of our universe. In applying the AdS/CFT correspondence to cosmology, it is assumed that each radial slice in the bulk direction corresponds to a time slice of the universe. Usually, it is assumed that the two sides of the brane have two different 5-dimensional bulk geometries ($\mathcal{M}_+$ and $\mathcal{M}_-$). The induced metrics on both sides of the brane must coincide to obtain a unique metric on the brane. This condition is called the first Israel junction condition, which fixes the tangential components of the two bulk metrics. Also, the first derivative of the metric on the brane is not continuous, this is due to the fact that the brane itself has its energy-momentum tensor. The requirement for the continuity of the first normal derivative of the metric across the brane leads to a constraint known as the second Israel junction condition. In the upcoming sections, we will see that this second Israel junction condition will be useful while deriving the time dependence of the brane radial coordinate.
	\\We start taking a four-dimensional FLRW universe on the brane which is embedded inside a five-dimensional bulk having bulk matter fields in it. The corresponding action reads \cite{Chakrabortty:2011sp,Park:2020jio}
	\begin{equation}
		S=\frac{1}{16\pi G}\int_{\mathcal{M}}d^{5}x \sqrt{-g}(\mathcal{R}-2\Lambda^{(\pm)})+S^{(\pm)}_{M}-\frac{1}{8\pi G} \int_{\partial \mathcal{M}}d^{4}x \sqrt{-\gamma}K^{(\pm)}
	\end{equation}
	In the above action, the first part corresponds to the bulk gravity action, the second part $S_M^{(\pm)}$ corresponds to the action of bulk matter fields, and the third one is the Gibbons-Hawking-York (GHY) boundary term \cite{York:1972sj,Gibbons:1976ue} which is required to make a well-defined variational problem. Here $\mathcal{R}$ is the Ricci scalar, $\Lambda^{(\pm)}$ is the cosmological constant  given by $-\frac{6}{R_{\pm}^2}$, $R_{\pm}$ is the AdS radius for $\mathcal{M_{\pm}}$ and $K$ is the trace of extrinsic curvature of $\partial\mathcal{M}$, where $\mathcal{M}$ is the bulk manifold. $\gamma$ is the determinent of the induced metric on the boundary.\\
	
	\begin{figure}[ht]
		\centering
		\includegraphics[width=0.9\linewidth]{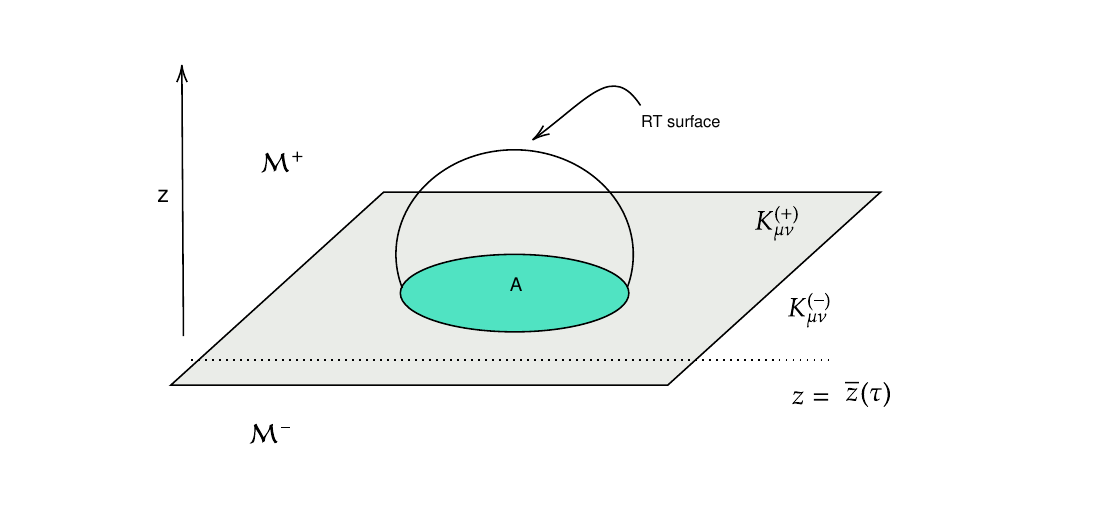}
		\caption{A schematic diagram of the braneworld model of cosmology along with a subsystem $A$ and the corresponding RT surface anchored with it. $\mathcal{M}^{\pm}$ are the two bulk manifolds on both sides of the brane and $\mathcal{K}_{\mu\nu}^{\pm}$ are the extrinsic curvature tensors corresponding to $\mathcal{M}^{\pm}$. The brane is moving along the $z$ direction, with the current position of the brane at $z=\z(\tau)$.}
		\label{fig:enter-label}
	\end{figure}
	\noindent Now the most general bulk metric that gives translational and rotational symmetries on the brane has the following form
	\begin{equation}\label{ds general}
		ds^{2}=-A(r)dt^{2} + B(r)dr^{2}+C(r)\delta_{ij}dx^{i}dx^{j}
	\end{equation}
	where $i,j=1,2,3$. It is very important to note that the radial range of the bulk is not extended up to infinity due to the existence of the brane. If we place the brane at $\Bar{r}$, then the bulk AdS space is extended in the range $0<r<\Bar{r}$. In order to construct the braneworld model, we have to consider another AdS space on the other side of the brane, which is extended in the range $\Bar{r}<r<\infty$. If we additionally impose the requirement of reflection invariance under the transformation $r\to 2\Bar{r}-r$, the allowed radial ranges for the two bulk spaces are constrained to $0<r<\Bar{r}$ and $\Bar{r}<r<2\Bar{r}$. Usually, the two sides of the brane have two different bulk geometries but for simplicity, we will consider a system having $\mathcal{Z}_{2}$ symmetry, hence we have to deal with only one kind of bulk throughout our calculations. In this case if the geometry of $\mathcal{M}_{-}$ is known the geometry of $\mathcal{M}_{+}$ is immediately fixed by the $\mathcal{Z}_{2}$ symmetry. \\
	Now we will proceed further to derive the second Israel junction condition. To derive this condition we have to convert the bulk spacetime metric using Arnowitt-Deser-Misner (ADM) decomposition \cite{Arnowitt:1959ah,Arnowitt:1962hi}. After using the equations of motions, this action will lead to an on-shell gravity action. Now taking the variation of the action with respect to the boundary metric, one gets \cite{Chamblin:1999ya}
	\begin{equation}\label{T brane pm}
		\delta S= -\frac{1}{16\pi G}\int_{\partial M}d^{4}x\sqrt{-\gamma}(K^{(\pm)}_{\mu\nu}-\gamma_{\mu\nu}K^{(\pm)})\delta \gamma^{\mu\nu}
	\end{equation}
	Hence, the stress-energy tensor of the boundary spacetime is given by
	\begin{equation}
		t^{(\pm)}_{\mu\nu}=-\frac{1}{16\pi G}(K^{(\pm)}_{\mu\nu}-\gamma_{\mu\nu}K^{(\pm)})~.
	\end{equation}
	Because of the $\mathcal{Z}_{2}$ symmetry, the extrinsic curvature tensor of the two bulk spacetimes satisfy $K^{(+)}_{\mu\nu}=-K^{(-)}_{\mu\nu}$. Therefore, if we denote $K_{\mu\nu}=K^{(-)}_{\mu\nu}$, the energy momentum tensor can be written as
	\begin{equation}\label{EMT pm}
		t^{(\pm)}_{\mu\nu}=\pm\frac{1}{16\pi G}(K_{\mu\nu}-\gamma_{\mu\nu}K)~.
	\end{equation}
	Now, due to the presence of the brane's non-vanishing stress tensor, the boundary stress tensors for the two bulk spacetimes ($\mathcal{M}^{\pm}$) are not the same. The discrepancy is addressed by the stress tensor of the brane, which is commonly referred to as the second Israel junction condition. This can be expressed mathematically as follows
	\begin{equation}\label{junction t}
		t^{(+)}_{\mu\nu}-t^{(-)}_{\mu\nu}=T_{\mu\nu}^{b}
	\end{equation}
	where $T_{\mu\nu}^{b}$ is the energy-momentum tensor on the brane, which is defined by
	\begin{equation}\label{T brane}
		T_{\mu\nu}^{b}=\frac{1}{\sqrt{-\gamma}}\frac{\delta S^{b}}{\delta\gamma_{\mu\nu}}.
	\end{equation}
	We would like to mention that in this analysis, the brane is in its ground state with constant energy density and pressure. In this case, the brane action is given by \cite{Chamblin:1999ya}
	\begin{equation}
		S^{b}=-\frac{\sigma}{4\pi G}\int_{\partial M}d^{4}x\sqrt{-\gamma}
	\end{equation}
	where $\frac{\sigma}{4\pi G}$ is the brane tension. Therefore, using the above expression of $S^{b}$ in eq.\eqref{T brane}, the stress tensor of the brane becomes 
	\begin{equation}\label{T brane gamma}
		T^{b}_{\mu\nu}=\frac{\sigma}{8\pi G}\gamma_{\mu\nu}. 
	\end{equation}
		Now using eq(s).(\eqref{EMT pm},\eqref{junction t}) along with eq.(\eqref{T brane gamma}), we get 
		\begin{equation}\label{jun initial}
			K_{\mu\nu}-\gamma_{\mu\nu}K=\sigma \gamma_{\mu\nu}~.
		\end{equation}
		In order to obtain the extrinsic curvature ($K$), we will multiply $\gamma^{\mu\nu}$ on the both sides of the above equation, which gives 
		\begin{equation}
			K=-\frac{4\sigma}{3}~.
		\end{equation}
		Substituting the expression of $K$ back in eq.(\eqref{jun initial}), we get
		the following expression for the extrinsic curvature tensor on the brane, that reads
	\begin{equation}
		K_{\mu\nu}=-\frac{\sigma}{3}\gamma_{\mu\nu}~.
	\end{equation}
	As we know in the braneworld model, the time-dependent cosmology on the brane is governed by the radial motion of the brane in the bulk direction. We will see that the above equation relating the extrinsic curvature tensor and induced metric on the brane will lead to an equation that will determine the time-dependent radial position of the brane. 
	\\To compute the extrinsic curvature on the brane we have to find the normal vector on the brane when it is moving in the radial direction. The unit normal vector for the spacetime metric in eq.\eqref{ds general} is given by   
	\begin{equation}
		n_{\alpha}=\frac{\sqrt{AB}}{\sqrt{A-B \dot r ^{2}}}\{\dot r,-1,0,0,0\}
	\end{equation}
	where $\dot r$ is the time derivative of the radial coordinate of the brane. We know that the extrinsic curvature tensor is given by 
	\begin{equation}
		K_{\mu\nu}=\gamma^{M}_{\mu}\gamma^{N}_{\nu}\nabla_{M}n_{N}~.
	\end{equation}
	Using the above equation, one can calculate the special component of the extrinsic curvature, which reads
	\begin{equation}
		K_{ij}=-\frac{\sqrt{AB}}{B}\frac{C'}{C}\frac{1}{\sqrt{A-B\dot r^{2}}}\gamma_{ij}
	\end{equation}
	where $\prime$ denotes the derivative with respect to the radial coordinate $r$.  Focusing on the special components of the extrinsic curvature and the induced metric, the second Israel junction condition simplifies to the  form
	\begin{equation}\label{C prime C}
		\frac{C'}{C}=\frac{\sigma}{3}\sqrt{\frac{B}{A}}\sqrt{A-B\dot r^{2}}~.
	\end{equation}
	The above equation contains the time derivative of the radial coordinate, however, in order to study the time dynamics of the universe we must introduce the cosmological time ($\tau$). To do so we will do the following reparametrization
	\begin{equation}\label{re para}
		-d\tau ^{2}=-A(r)dt^2 +B(r)dr^2~.
	\end{equation}
	Under this re-parametrisation, the spacetime metric in eq.\eqref{ds general} becomes the induced metric on the brane. Now identifying $C(r)$ as the scale factor of the universe ($a(\tau)$) the induced metric on the brane becomes
	\begin{equation}
		ds_{b}^{2}=-d\tau^{2}+a(\tau)^2 \delta_{ij}dx^i dx^j~.
	\end{equation}
	The induced metric is essentially the FLRW metric, which describes the time evolution of the universe. This is why the parameter $\tau$ is referred to as the cosmological time. \\
	Finally, we derive the junction condition. To do so we will start with eq.(\eqref{C prime C}). Upon rearranging eq.(\eqref{C prime C}), we can obtain the following expression of $\dot{r}$
	\begin{equation}
		\dot{r}^{2}=\frac{A}{B}-\frac{9A}{\sigma^2 B^2}\frac{C^{\prime 2}}{C^2}~.
	\end{equation}
	Now we will use eq.\eqref{re para} to express the above equation in terms of the derivative of the radial coordinate with respect to cosmological time, which finally gives 
	\begin{equation}\label{Israel junction final}
		\left(\frac{dr}{d\tau}\right)^2 =\left(\frac{\sigma}{3}\right)^2 \left(\frac{C}{C'}\right)^2 -\frac{1}{B}~.
	\end{equation}
	Later, this equation will be useful to determine the time-dependent radial position of the brane for different matter-dominated universes.
	\section{Cosmological $p$-brane gas geometry model}\label{Cosmological p-brane gas geometry model}
	In the previous section, we have shown the dependence of the radial motion of the brane with the cosmological time. 
	In this section, we will provide a brief introduction to the cosmological p-brane gas geometry model. The action governing the p-branes distributed in the $(d+1)$-dimensional AdS space is given by \cite{Park:2021wep}
	\begin{equation}
		S=\frac{1}{16\pi G}\int d^{d+1}x \sqrt{-g}(\mathcal{R}-2\Lambda)+T_{p}N_{p}\int d^{p+1}\xi \sqrt{-h}\partial^{\alpha}x_{\mu}h_{\alpha\beta}\partial^{\beta}x_{\nu}g^{\mu\nu}
	\end{equation}
	where $N_{p}$ and $T_{p}$ are number and tension of p-branes respectively. Here $x^{M}(M=0,....,d)$ and $\xi^{\alpha}(\alpha =0,...,p-1,d)$ are the coordinates of the bulk spacetime and the brane world volume. Now the variation of the action with respect to the bulk metric reads
	\begin{equation}
		\delta S =\frac{1}{16\pi G}\int d^{d+1}x \sqrt{-g}\left(\mathcal{R}_{\mu\nu}-\frac{1}{2}\mathcal{R} g_{\mu\nu}+\Lambda g_{\mu\nu}\right)\delta g^{\mu\nu}+T_{p}N_{p}\int d^{p+1}\xi \sqrt{-h}\partial^{\alpha}x_{\mu}h_{\alpha\beta}\partial^{\beta}x_{\nu}\delta g^{\mu\nu}
	\end{equation}
	We would like to mention that in the above step while varying the action we have not written the GHY boundary term as this will not affect the equation of motion. Before proceeding further to compute the equation of motion one should note that the integral over the bulk and the brane world volume has different dimensions, hence it is problematic to write the Einstein's equations directly. To resolve this problem, one has to assume that p-branes are uniformly distributed in spatial directions perpendicular
	to the brane’s world volume. If we denote 
	\begin{equation}
		T_{MM}=-\frac{T_{p}n_{p}R^{d-p}}{r^{d-p}}\left\{g_{tt},\frac{p-1}{d-1}g_{11},\frac{p-1}{d-1}g_{22},\dots,\frac{p-1}{d-1}g_{(d-1)(d-1)},g_{rr}\right\}
	\end{equation}
	One can derive Einstein's equations as follows
	\begin{equation}
		R_{MN}-\frac{1}{2}g_{MN}R+\Lambda g_{MN}=16\pi G T_{MN}
	\end{equation}
	Now to solve Einstein's equation let us take the ansatz for the spacetime metric to be AdS black brane\footnote{Due to the presence of a negative cosmological constant the most general spacetime geometry will be an AdS black brane.}, that is
	\begin{equation}\label{BB metric}
		ds^{2}=\frac{r^2}{R^2}\left(-f(r)dt^2 + \delta_{ij}dx^i dx^j\right)+\frac{R^2}{r^2 f(r)}
	\end{equation}
	where, $i,j =2,3\dots,d$. This ansatz along with Einstein's equation leads to the following expression of the brane's lapse function \cite{Park:2021wep}
	\begin{equation}
		f(r)=1-\frac{\rho_{p}}{r^{d-p}}
	\end{equation}
	where $\rho_{p}$ is the energy density of the p-brane gas.  
	\begin{equation}
		\rho_p =8\pi G c_{p}n_{p}T_{p}R^{d-p+2}
	\end{equation}
	where $c_p$ is constant whose value is fixed by the number $p$. As we want to study a (3+1)-dimensional FLRW universe on the brane, we must consider a (4+1)-dimensional bulk spacetime. Hence the lapse function becomes $f(r)=1-\frac{\rho_p}{r^{4-p}}$.For different values of $p$, different kinds of matter sources will be present in the braneworld model. In the p-brane gas model, the scale factor of standard cosmology is influenced by the radial position of the brane. By treating p-branes as Dp-branes within the framework of string theory, the authors in \cite{Park:2021wep} derived an expression for the scale factor in the braneworld scenario, which is given as follows
	\begin{equation}
		a(\tau)\sim \tau^{\frac{2}{4-p}}~.
	\end{equation}
	By comparing this result with the eq. , we can derive the relationship between $p$ and the equation of state parameter $\omega$, which is expressed as follows
	\begin{equation}\label{omega p rel}
		p=4-3(1+\omega)~.
	\end{equation}
	Based on our understanding of standard cosmology, we know various values of the equations of state parameters. This, in turn, allows us to identify the corresponding p-brane configurations in the bulk that give rise to the different types of matter present in the braneworld model. Below we will find different values of $p$ corresponding to three different matter sources in the braneworld model
	\begin{enumerate}
		\item In the presence of no matter fields in the braneworld model, we do not consider any p-brane gas in the bulk. In this case, the bulk geometry is a pure AdS spacetime. Hence from eq.\eqref{Israel junction final}, we obtain
		\begin{equation}
			\paren{\frac{dr}{d\tau}}^2=\paren{\frac{\sigma^2}{36}-\frac{1}{R^2}}r^2~.
		\end{equation}
		This implies the radial position of the brane is given by
		\begin{equation}
			\Bar{r}(\tau)=r_{0}e^{H\tau}
		\end{equation}
		Where $r_0$ is the initial position of the brane.
		\item In a four-dimensional FLRW universe, $\omega =\frac{1}{3}$, indicates a radiation-dominated era. Hence, from eq.\eqref{omega p rel} one can find $p=0$. This means that a bulk 0-brane is dual to radiation on the brane and the scale factor in the radiation-dominated era increases by $a(\tau)\sim \tau^{1/2}$. In the radiation-dominant era, the lapse function of the black brane geometry is given by
		\begin{equation}\label{lapse rad}
			f(r)=1-\frac{m}{r^4}
		\end{equation}
		where $m$ is the energy density of 0-branes. From the AdS/CFT point of view, a black hole solution in AdS space is dual to a thermal system of
		a gauge theory with finite temperature. It should also be mentioned that the parameter $m$ acts as the excitation energy of gauge bosons on the dual field theory side. The corresponding Isreal junction condition can be obtained from eq.\eqref{Israel junction final}, which reads
		\begin{equation}
			\paren{\frac{dr}{d\tau}}^2 =\paren{\frac{\sigma^2}{36}-\frac{1}{R^2}}r^2 +\frac{m}{R^2}\frac{1}{r^2}
		\end{equation}
		To only consider the non-trivial contribution from the radiation term we need to solve the above equation for $\sigma=\sigma_{c}$, which leads to the following differential equation
		\begin{equation}
			\frac{dr}{d\tau}=\frac{\sqrt{m}}{R}\frac{1}{r}
		\end{equation}
		In the braneworld model, the radial position of the brane corresponds to the scale factor in brane cosmology. As a result, solving the differential equation yields the radial position of the brane as a function of cosmological time. That is given by
		\begin{equation}
			\Bar{r}(\tau)=\frac{\sqrt{2}m^{1/4}}{\sqrt{R}}\tau^{1/2}+r_i
		\end{equation}
		where $r_i$ is the initial position of the brane. This means $\Bar{r}(\tau =0)=r_i$.
		\item  In a matter-dominated era, the equation of state parameter $\omega=0$, hence $p=1$, indicates that the backreaction of a 1-brane in bulk acts as a source of matter fields in the braneworld. For a matter-dominated era, the lapse function becomes 
		\begin{equation}
			f(r)=1-\frac{\text{\huge}\rc}{r^3}
		\end{equation} 
		The corresponding Isreal junction condition can be obtained from eq.\eqref{Israel junction final}, which reads
		\begin{equation}
			\paren{\frac{dr}{d\tau}}^2 =\paren{\frac{\sigma^2}{36}-\frac{1}{R^2}}r^2 +\frac{\rc}{R^2}\frac{1}{r}
		\end{equation}
		Similar to the radiation case if we have to observe the non-trivial contribution of the brane position due to the presence of matter in the braneworld model, one needs to consider $\sigma =\sigma_{c}$. This gives
		\begin{equation}
			\frac{dr}{d\tau}=\frac{\sqrt{\rc}}{R}\frac{1}{r^{1/2}}~.
		\end{equation}
		Now solving the above differential equation, we get the radial position of the brane as follows
		\begin{equation}
			\Bar{r}(\tau)=\paren{\frac{3}{2}}^{2/3}\frac{\rc^{1/3}}{R^{2/3}}+r_i
		\end{equation}
		\item In the previous two points, we have discussed about ordinary matter sources. Now we will also take some exotic matter for which the equation of state parameter ($\omega$) is $-\frac{1}{3}$. The reason behind calling them exotic matter is that these kinds of matter have negative mass or energy density. Although they are useful in the context of dark energy, dark matter and linear expansion of the universe. The blackening factor corresponding to this kind of matter is given by 
		\begin{equation}
			f(r)=1-\frac{\delta}{r^2}
		\end{equation}
		The corresponding Israel junction condition can be obtained from eq.\eqref{Israel junction final}, which reads
		\begin{equation}
			\paren{\frac{dr}{d\tau}}^2 =\paren{\frac{\sigma^2}{36}-\frac{1}{R^2}}r^2 +\frac{\delta}{R^2}
		\end{equation}
		For $\sigma=\sigma_{c}$, the above differential equation reduces to
		\begin{equation}
			\frac{dr}{d\tau}=\frac{\delta^{1/2}}{R}
		\end{equation}
		Solving the above equation we get 
		\begin{equation}\label{bar r ex}
			\Bar{r}=\frac{\sqrt{\delta}}{R}\tau +r_{i}
		\end{equation}
		where $r_{i}$ is the initial position of the brane.
		From the above expressions, it is clear that the blackening factors for radiation, matter and cosmic strings have different power behaviours. This is due to the fact that for the first kind of solution, matter should be localized in the radial direction, and for the second kind one-dimensional objects should be extended along the radial direction and when cosmic strings are concerned two-dimensional objects are extended along the radial direction. Because of the different dimensions of the objects along the radial direction, the blackening factors have different forms.\\
		It is obvious to ask what happens to stiff matter in the braneworld model of cosmology. From the results of standard cosmology, it is well known that the equation of state parameter, $\omega=1$. Hence, from eq.\eqref{omega p rel} we can find $p=-2$. This is unphysical as $p$ denotes the dimension of the p-brane gas in the bulk spacetime of the braneworld model. Previous studies have mentioned that stiff matter was only present in the early universe when the energy scale of our universe was very high. Hence it may be possible that the effects of stiff matter could be observed in higher dimensions only. That is why the presence of the stiff matter seems unphysical in the $(4+1)$-dimensional braneworld model. We leave further discussions on this point for future work.
	\end{enumerate}
	\section{Entanglement entropy of the universe}\label{Entanglement entropy of the universe}
	In this section, we will calculate time-dependent entanglement entropy for various matter-dominated universes and also show their early and late time behaviours. We have computed the results holographically for a small spherical subsystem on the brane. 
	\subsection{Time independent entanglement entropy for pure AdS spacetime}\label{TIE AdS}
	In this section, we will study the time-independent entanglement entropy of the four-dimensional static FLRW universe in the absence of any matter source. In the braneworld model, to calculate the HEE and HSC, we have to consider pure $AdS_5$ spacetime in the bulk. In order to do so we will consider a static brane for which $\sigma =\sigma_{c}$. However, we should remember that this model does not explain the expansion of the universe, but this is useful for understanding the time evolution of HEE and HSC in various expanding universes. 
	The spacetime metric for AdS$_5$ spacetime in the absence of any matter source in the universe is given by 
	\begin{equation}\label{AdS metric}
		ds^2 = \frac{R^2}{z^2}\paren{dz^2 -dt^2 +du^2 +u^2 d\Omega^2_{2}}~.
	\end{equation}
	According to the RT formula\cite{Ryu:2006ef,Ryu:2006bv}, the HEE for a subsystem $A$ on the boundary is given by
	\begin{equation}
		S_{HEE}=\frac{1}{4G^{(d+1)}_{N}}[Area(\Gamma^{A}_{min})]~.
	\end{equation}
	where $G^{(d+1)}_{N}$ is the Newton’s gravitational constant in $(d + 1)$-spacetime dimensions and $\Gamma^{A}_{min}$ is the $(d-1)$-dimensional static minimal surface such that $\partial \Gamma^{A}_{min}=\partial A$. We will now proceed to calculate the HEE for a circular subsystem on the brane. To do so we will divide the brane into two regions, that is, $0\leq u\leq l$ and $l\leq u< \infty$ with a two-sphere of radius $l$. To calculate the RT area functional we will consider a constant time slice, this means that we will set $dt=0$. Then we will parameterize the bulk coordinate in terms of a coordinate on the brane, that is, $z=z(u)$. Considering this parameterization, the RT area functional for the metric given in eq.\eqref{AdS metric} reads
	\begin{equation}\label{area func AdS}
		A^{(AdS)}=\Omega_{2}\int_{0}^{l}du\frac{\sqrt{1+z^{\prime 2}}}{z^3}~.
	\end{equation}
	We identify the integrand of the above area functional to be a Lagrangian of the form $\mathcal{L}=\mathcal{L}(z,z^\prime)$. Hence the Euler-Lagrange equation of motion corresponding to this Lagrangian reads
	\begin{equation}
		3u + 3u z^{\prime 2}(u)+ u z(u) z^{\prime\prime}(u)+2z(u) z^{\prime}(u)+ 2 z(u) z^{\prime 3}(u)=0~.
	\end{equation}
	To evaluate the form of the RT surface, we need to solve the above differential equation. However, it seems a bit difficult to solve the above equation as it is a non-linear differential equation. After doing a careful analysis,  we can write the above equation in the following form
	\begin{equation}
		u \frac{d}{du}(z(u)z^{\prime}(u)+u) + 2(1+z^{\prime 2}(u))(z(u)z^{\prime}(u)+u)=0~.
	\end{equation}
	It is easy to see that the above equation has a trivial solution of the form $z(u)z^{\prime}(u)+r=0$. Hence the solution of $z(u)$ is given by
	\begin{equation}
		z(u)=\sqrt{c_1 -(c_2 + u)^2}
	\end{equation}
	where $c_1$ and $c_2$ are arbitrary integration constants. To determine the value of $c_1$, $c_2$ we need to consider the boundary conditions for $z(u)$. The smoothness condition of the RT surface leads to $z^{\prime}(0)=0$ due to the rotational symmetry. Now this condition fixes the integration constant $c_2 =0$. Another condition that we invoke is that $z(l)=\z$, where $\z$ is the position of the brane. This condition leads to 
	\begin{equation}
		c_1 =l^2 +\z^2~.
	\end{equation}
	Therefore, the final form of the RT surface has the form
	\begin{equation}\label{z0(u)}
		z(u)=\sqrt{l^2 +\z^2 -u^2}~.
	\end{equation}
	We would like to mention that inside the subsystem, on the brane, $l^2 +\z^2$ is always much less than $u^2$. Hence for simplicity, while evaluating the area functional in eq.\eqref{area func AdS}, we will expand the integrand in power series of $\frac{u^2}{\C}$ and keep terms up to $\mathcal{O}(\frac{u^4}{\C^2})$. With the above consideration, substituting the expression of $z(u)$ from eq.\eqref{z0(u)} in eq.\eqref{area func AdS}, we get
	\begin{equation}
		A^{(AdS)}=\Omega_{2}\left[\frac{l}{2\sqrt{\C}\z^2}+\frac{1}{2\C}tan^{-1}\paren{\frac{l}{\sqrt{\C}}}\right].
	\end{equation}
	As $l<\sqrt{\C}$ ,we can expand the second term of the above expression. Keeping the first term of the expansion, we get
	\begin{equation}
		A^{(AdS)}\approx\Omega_{2}\left[\frac{l}{2\sqrt{\C}\z^2}+\frac{l}{2\C^{3/2}}\right]~.
	\end{equation}
	According to the RT formula, the HEE will therefore be
	\begin{equation}\label{HEE AdS}
		S^{(AdS)}_{HEE}=\frac{A^{(AdS)}}{4G}=\frac{\Omega_{2}}{4G}\left[\frac{l}{2\sqrt{\C}\z^2}+\frac{l}{2\C^{3/2}}\right]~.
	\end{equation}
	\subsection{Time-dependent entanglement entropy in an eternally inflating universe}
	In order to calculate the HEE of an eternally inflating universe, we will consider a non-static brane with $\sigma \neq \sigma_c$. In this case also, the bulk geometry is given by pure AdS spacetime. The main difference from the previous case is that here the brane moves with a velocity $\frac {dr}{d\tau}=\sqrt{(\frac{\sigma^2}{36}-\frac{1}{R^2})}r$ in the bulk direction. The calculation of HEE is not much affected by the radial velocity of the brane. But now the brane's position is a function of cosmological time ($\tau$). Similar to the previous case we will take the profile of the RT surface as 
	\begin{equation}
		z(u,\tau)=\sqrt{l^2 +\z(\tau)^2 -u^2}~.
	\end{equation}
	It should be mentioned that the time dependence of the entanglement entropy in the braneworld model arises from $\z(\tau)$. Now due to this time dependence of the brane's radial position, the properties of the HEE in the braneworld model change significantly. \\We will start by considering a circular subsystem of size $l$, which is measured in the comoving frame. Now due to the nontrivial scale factor of the brane, the actual size of the subsystem is given by
	\begin{equation}
		L=\frac{R}{\z(\tau)}l~.
	\end{equation}
	It should be mentioned that although the subsystem has time-independent finite size $l$, the physical subsystem size $L$ changes with time because our universe is expanding.  From the Isreal junction conditions, we have seen that the scale factor of the brane ($\sim \frac{1}{\z(\tau)}$) changes due to the presence of different matter sources, hence time dependence of HEE will also be different for different matter sources. In the braneworld model, the radial position of the brane is referred to as the scale factor of the universe and depends upon the cosmological time. Hence, the radial position of the brane is given by
	\begin{equation}
		\z(\tau)=\frac{R^2}{r_{0}}e^{-H\tau}~.
	\end{equation}
	We will now proceed to compute the HEE of the universe in the absence of any kind of matter. First, we will consider the early time limit, which tells $H\tau<<1$. In this case, the brane is approximately given by
	\begin{equation}\label{z bar eter early}
		\z(\tau)=\frac{1}{r_{0}}(1-H\tau).
	\end{equation}
	Now substituting this expression of $\z(\tau)$ in eq.\eqref{HEE AdS}, we get the early time behaviour for the HEE of the universe, which reads \footnote{We have kept the AdS radius $R=1$ for simplicity. }
	\begin{equation}
		S^{(AdS)}_{HEE}=\frac{\Omega_2}{4G}\left[\frac{lr_{0}^3 (1+2H\tau)}{\sqrt{r_{0}^2 l^2+1}}\paren{1+\frac{H\tau}{r_{0}^2 l^2+1}}+\frac{lr_{0}^3}{2(r_{0}^2 l^2+1)^{3/2}}\paren{1+\frac{3H\tau}{r_{0}^2 l^2+1}}\right]~.
	\end{equation}
	From the above equation, it is clear that the leading order behaviour of the HEE in the early universe increases linearly with cosmological time. One can also notice that for the early time of eternally inflating universe, HEE and brane's position scales in a similar fashion with the cosmological time.\\
	Now we will proceed to compute the late time behavior of HEE for an eternally inflating universe. For late time, the brane position is given by
	\begin{equation}\label{z bar eter late}
		\z(\tau)\approx\frac{e^{-H\tau}}{r_0}~.
	\end{equation}
	Using this expression of $\z(\tau)$, we can get the late time behaviour of the HEE. This is given by
	\begin{equation}
		S^{(AdS)}_{HEE}=\frac{\Omega_2}{4G}\left[\frac{r_{0}^2 e^{2H\tau}}{2}\paren{1-\frac{e^{-2H\tau}}{2l^2 r_{0}^2}}+\frac{1}{2l^2}\paren{1-\frac{3e^{-2H\tau}}{l^2 r_{0}^2}}\right]~.
	\end{equation}
	Hence, we can say that in the late time, the leading order of the HEE for an eternally inflating universe changes as $e^{2H\tau}$. It should be noted that in the late time era of an eternally inflating universe, the actual length scale ($L$) changes as $e^{H\tau}$. Therefore, based on the area law of HEE, one would anticipate that it should scale as $L^2$ at late times in an eternally inflating universe. From our expression for HEE during the late-time era, it is evident that its leading behaviour increases as $e^{2H \tau}$, which aligns with our expectations.
	\subsection{Holographic entanglement entropy in the presence of radiation}
\begin{figure}[ht!]
		\centering
		\includegraphics[width=0.7\linewidth]{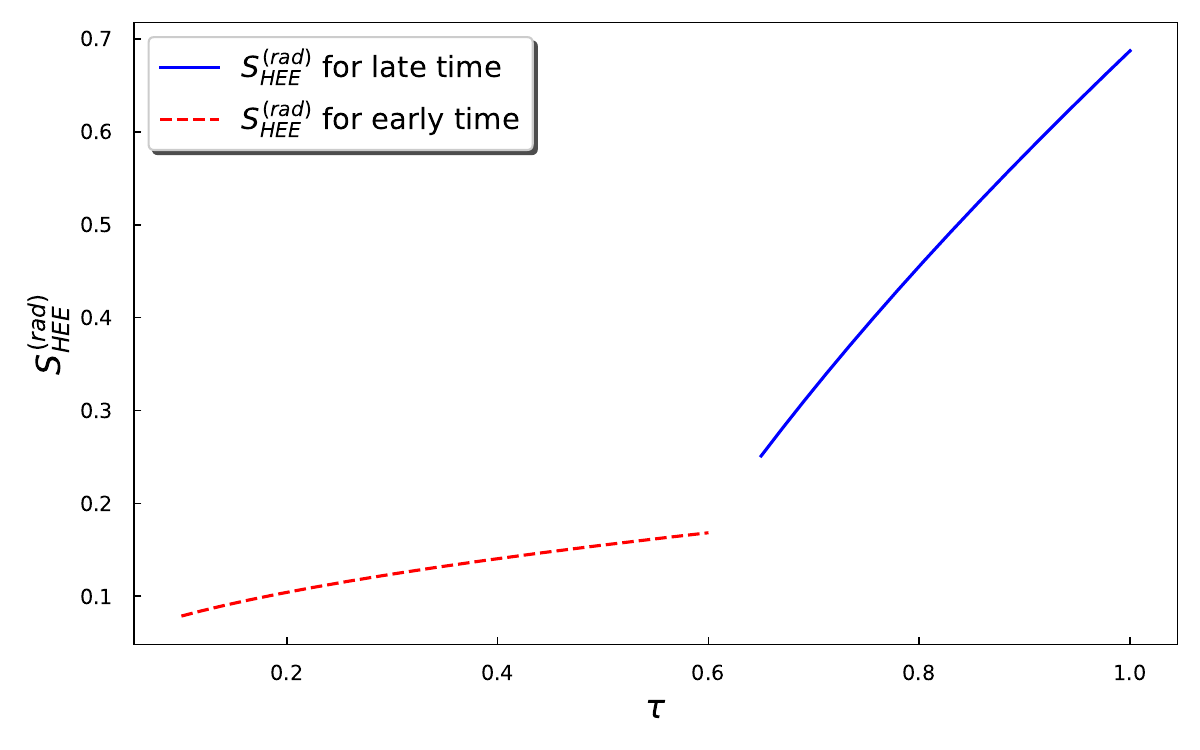}
		\caption{The above plot shows the variation of holographic entanglement entropy with cosmological time for the early and late time regimes of the radiation-dominated universe. This plot is done for $z_h =2$, $l=2$, $\T{m}=0.001$ and $z_i =8$. The dashed graph in red refers to the early time behaviour, and the solid line in blue refers to the late time behaviour of radiation dominated universe. Here, the prefactor $\frac{\Omega_2}{4G}$ is set to 1.}
		\label{fig:Srad}
	\end{figure}

	In this section, we will calculate the entanglement entropy of the $(3+1)$-dimensional FLRW universe in the presence of radiation holographically. We will begin by examining a $(4+1)$-dimensional bulk spacetime characterised by a black brane geometry. This spacetime has a lapse function defined as specified in. \eqref{lapse rad}. Hence, considering the spacetime metric of the black brane in eq.\eqref{BB metric} in the  $z$ coordinate and following the same procedure as  in the previous section, we can compute the area functional as follows 
	\begin{equation}\label{Area func rad}
		A^{(rad)}=\Omega_{2}\int_{0}^{l} du \frac{u^{2}}{z(u)^{3}}\sqrt{1+\frac{z^{\prime 2}(u)}{f(z(u))}}
	\end{equation}
	where $f(z(u))=1-\m z(u)^4$ and $\T{m}=\frac{m}{R^8}$.\\
	Now identifying the integrand as the  Lagrangian, we can compute the Euler-Lagrange equation of motion, which reads
	\begin{align}\label{eq z rad}  
		& 3 \T{m}^2 u z(u)^8 + 3 u \left( 1 + (z'(u))^2 \right) 
		- \T{m} u z(u)^4 \left( 6 + (z'(u))^2 \right) 
		- \T{m} z(u)^5 \left( 2 z'(u) + u z''(u) \right)\nonumber \\&
		+ z(u) \left( 2 \left( z'(u) + (z'(u))^3 \right) + u z''(u) \right)=0~.
	\end{align}
	To solve the above differential equation, we will assume a perturbative form for $z(u)$, which is given by
	\begin{equation}
		z(u)=z_{0}(u)+\m z_{1}(u)+\mathcal{O}(\frac{1}{z_{h}^8})~.
	\end{equation}
	Taking into account the perturbative form of \( z(u) \) mentioned earlier, the expression for the RT area functional can be expanded up to \( \mathcal{O}(\m) \) as follows
	\begin{align}\label{approx area func rad}
		A^{(rad)}&\approx \Omega_{2}\left[\frac{u^2}{z_{0}(u)^3}\sqrt{1+z^{\prime 2}_{0}(u)}\nonumber\right.\\&\left.+\m\frac{u^2}{z_{0}(u)^3}\sqrt{1+z^{\prime 2}_{0}(u)}\paren{-3\frac{z_{1}(u)}{z_{0}(u)}+\frac{2z^{\prime}_{0}(u)z^{\prime}_{1}(u)+z^{\prime 2}_{0}(u)z^{4}_{0}(u)}{2(1+z^{\prime 2}_{0}(u))}}\right]~.
	\end{align}
	Later, we will use the above approximated expression of $A^{(rad)}$ to evaluate the final form of the HEE.
	Now substituting the form of $z(u)$ in eq.\eqref{eq z rad}, we get two differential equations corresponding to $\mathcal{O}(\m^0)$ and $\mathcal{O}(\m^1)$, which reads
	\begin{equation}
		3 u + 2 z_0(u) z_0'(u) + 3 u (z_0'(u))^2 + 
		2 z_0(u) (z_0'(u))^3 + u z_0(u) z_0''(u)=0
	\end{equation}
	\begin{align}\label{z1 eq rad}
		& u z_1(u) z_0''(u) + u z_0(u) z_1''(u)- 
		u z_0(u)^5 z_0''(u)-6 u z_0(u)^4 - 2 z_0(u)^5 z_0'(u) \nonumber\\&+ 
		2 z_1(u) z_0'(u) - u z_0(u)^4 (z_0'(u))^2 + 
		2 z_1(u) (z_0'(u))^3 + 2 z_0(u) z_1'(u) \nonumber\\&+ 
		6 u z_0'(u) z_1'(u) +6 z_0(u) (z_0'(u))^2 z_1'(u) =0~.
	\end{align}
	From the above two equations, it is clear that the first one corresponds to the minimal RT surface of pure AdS spacetime. The solution of this equation is already obtained in section \eqref{TIE AdS} and has the following form
	\begin{equation}
		z_{0}(u)=\sqrt{l^2 +\z^2 -u^2}~.
	\end{equation}
	Now in order to solve the eq.\eqref{z1 eq rad}, we will substitute the above expression of $z_{0}(u)$ in eq.\eqref{z1 eq rad}. Also to do a series solution, we will expand the terms of eq.\eqref{z1 eq rad} up to $\mathcal{O}(\frac{u^4}{\C^2})$. If we take the first three terms of the series solution of eq.\eqref{z1 eq rad} , then the expression of $z_{1}(u)$ becomes
	\begin{equation}
		\begin{split}
			z_{1}(u)&=a_{0}\paren{1+\frac{u^2}{2\C}+\frac{3u^4}{8\C^2}+\mathcal{O}\paren{\frac{u^6}{\C^3}}}\\ &+\C^{5/2}\left(\frac{u^2}{\C}\right)\left(\frac{1}{2}-\frac{13}{40}\frac{u^2}{\C}+\frac{23}{560}\frac{u^4}{\C^2}\right. \\& \left.+\mathcal{O}\paren{\frac{u^5}{\C^{5/2}}}\right)
		\end{split}
	\end{equation}
	where $a_{0}$ is an arbitrary constant, which will be determined by the boundary condition of $z_{1}(u)$. From the boundary condition of $z_{0}(u)$, we can say that $z_{1}(l)=0$ as $z_{0}(l)=\z$. This boundary condition gives 
	\begin{equation}\label{z1 rad}
		z_{1}(u)=\C^{5/2}\paren{\frac{u^2}{2\C}-\frac{l^2}{2\C}-\frac{13}{40}\frac{u^4}{\C^2}-\frac{1}{4}\frac{l^2u^2}{\C^2}+\frac{23}{40}\frac{l^4}{\C^2}}~.
	\end{equation}
	Now using eqs.(\eqref{z0(u)},\eqref{z1 rad}) in eq.\eqref{approx area func rad}, we can evaluate the expression of $A^{(rad)}$, which is given by
	\begin{equation}
		A^{(rad)}=\Omega_{2}\left[\frac{l}{2\sqrt{\C}\z^2}+\frac{l}{2\C^{3/2}}+\m \frac{l^5}{10 (l^2 + \z ^{2})^2}\right]~.
	\end{equation}
	We would like to mention that while evaluating the integral in eq.\eqref{approx area func rad}, the second term in the integrand is expanded up to $\mathcal{O}(\frac{u^4}{\C^2})$ before performing the integral.
	The first term in the area functional represents the contribution from pure AdS geometry, while the second term serves as a correction to this pure AdS term, accounting for the effects of radiation within the braneworld model.\\
	Hence, from the RT proposal, the HEE is given by \cite{Ryu:2006bv}
	\begin{equation}
		S^{(rad)}_{HEE}=\frac{\Omega_{2}}{4G}\left[\frac{l}{2\sqrt{\C}\z^2}+\frac{l}{2\C^{3/2}}+\m \frac{l^5}{10 (l^2 + \z ^{2})^2}\right]~.
	\end{equation}
	With the above expression of HEE for the radiation-dominated era in hand, we will now proceed to calculate the early and late time behaviour of it. \\From the Israel junction condition of the radiation-dominated era, we get  
	\begin{equation}\label{rad z bar}
		\z =\frac{z_{h}z_{i}}{z_{h}+\sqrt{2}z_{i}\sqrt{\tau}}~.
	\end{equation}
	For early time $z_{h}\sqrt{R}>z_{i}\sqrt{\tau}$, hence $\z$ can be approximated as follows
	\begin{equation}\label{z bar rad early}
		\z \approx (1-\frac{\sqrt{2}z_{i}}{z_{h}}\sqrt{\tau})z_{i}~.
	\end{equation}
	Now using the above expression of $\z$, we get the following expression for $S^{(rad)}_{HEE}$
	\begin{align}\label{srad early time}
		S^{(rad)}_{HEE}=\frac{\Omega_2}{4G}\Bigg[\frac{l}{2z_{i}^2 \sqrt{\ci}}\paren{1+\eta_1 \sqrt{\tau}}\paren{1+\frac{2\sqrt{2}z_{i}}{z_h}\sqrt{\tau}}+\frac{\m l^5 \paren{1+4\eta_{1}\sqrt{\tau}}}{10 \ci^2}\Bigg]~
	\end{align}
	where $\eta_1 =\frac{\sqrt{2}z_i^3}{z_h \ci}$. From the above expression, it is clear that in the early time of the radiation-dominated universe, the HEE scales as $S^{(rad)}_{HEE}\sim \tau^{1/2}$. We would like to mention that in the early time of the radiation-dominated universe, the brane position and HEE both scale as $\tau^{1/2}$. Hence, in the early time, the brane's radial velocity and rate of change of HEE with the cosmological time have similar scaling behaviour with $\tau$. \\We will move forward to compute the late time behaviour of HEE in the late time of the radiation-dominated universe. In the late time era ($\tau \to \infty$), the radial position of the brane is approximately given by 
	\begin{equation}
		\z \approx\frac{z_{h}}{\sqrt{2\tau}}~.
	\end{equation}
	With this approximate value of the brane's position, the late-time behaviour of the HEE in a radiation-dominated era reads 
	\begin{equation}\label{srad late time}
		S^{(rad)}_{HEE}=\frac{\Omega_2}{4G}\left[\frac{\tau}{z_{h}^2}\left(1-\frac{z_{h}^2}{4\tau l^2}\right)+\frac{1}{2l^2}\left(1-\frac{3 z_{h}^2}{4\tau l^2}\right)+\m\frac{l^4}{10}\left(1-\frac{z_{h}^2}{\tau l^2}\right)\right]~.
	\end{equation}
	The above equation suggests that the leading order behaviour of the HEE with cosmological time is given by $S^{(rad)}_{HEE}\sim \tau$.  In the late time of radiation dominated era, the physical length scale changes as $L\sim \sqrt{\tau}$. Therefore, according to the area law, the HEE should change as $L^2$. From the above equation, we can see that the leading order behaviour of $S^{(rad)}_{HEE}$ in the late time era scales as $\tau$. Hence, our expression of $S^{(rad)}_{HEE}$ is consistent with the well known area law of entanglement.\\
	We plot our results in Fig.\eqref{fig:Srad}. The dashed red line is the graphical representation of eq.\eqref{srad early time} and the solid line in blue refers to the graphical representation of eq.\eqref{srad late time}. These plots have been done for the parameters $z_h =2$, $l=2$, $\T{m}=0.001$ and $z_i =8$. We would like to mention that the parameters ($z_h$, $l$, $\T{m}$ and $z_i$) are chosen such that they obey the bound $z_{h}\sqrt{R}>z_{i}\sqrt{\tau}$. From the plot, it is evident that after a certain time, the late-time expression of HEE for a radiation-dominated universe dominates its early-time behaviour.
	\subsection{Holographic entanglement entropy in the presence of matter}
	In this section, we will calculate the time-dependent entanglement entropy in the presence of matter in the braneworld model. Similar to the radiation case we will parameterise $z=z(u)$ and set $dt=0$ in eq.\eqref{BB metric}. This will allow us to compute the area functional corresponding to the black brane geometry in the bulk and have a blackening factor $f_m=1-\T{\rc} z(u)^3$, with $\T{\rc}=\frac{\rc}{R^6}$. The area functional reads
	\begin{equation}\label{Area mat}
		A^{(mat)}=\Omega_{2}\int_0^ldu\frac{u^2}{z(u)^3}\sqrt{1+\frac{z'^2}{f_m\left(z(u)\right)}}~.
	\end{equation}
	Now identifying the integrand as the Lagrangian, we can compute the Euler-Lagrangian equation for $z(u)$, which reads
	\begin{align}\label{z eq mat}
		& 6\rc^2u z^6(u)+6u\paren{1+z^{'2}(u)}-3\rc u z^3(u)\paren{4+z^{'2}(u)}-2\rc z^4(u)\paren{2z^{'}(u)+uz^{''}(u}\\ &+2z(u) \paren{2\paren{z^{'}(u)+z^{'}(u)^3}+u z^{''}(u}=0~.
	\end{align}
	To solve the differential equation, we will once again consider a perturbative representation of the profile of the RT surface. For small subsystem size $z_{0}^3 \T{\rc}<<1$, the profile of the RT surface is given by the following perturbative form
	\begin{equation}
		z(u)=z_{0}(u)+\T{\rc}z_{1}(u)+\mathcal{O}(\T{\rc}^2)~.
	\end{equation}
	Again substituting the above expression of $z(u)$ in eq.\eqref{z eq mat}, we get the following two equations corresponding to $\mathcal{O}(\rc^0)$ and $\mathcal{O}(\rc^1)$
	\begin{equation}
		\left( 3 u + 2 z_0(u) z_0'(u) \right) \left( 1 + (z_0'^2(u) \right) + 
		u z_0(u) z_0''(u)=0
	\end{equation}
	\begin{align}
		& 2z_0(u)\paren{\paren{2+6z_0'^{2}(u)}z_1'(u)+uz_1''(u)}-2z_0^4(u)\paren{2z_0'(u)+uz_0''(u)}+2uz_1(u)z_0''(u)\nonumber \\&+4z_0'(u)\paren{z_1(u)\paren{1+z_0^{'2}(u)}+3uz_1^{'}(u)}-3uz_0^3(u)\paren{4+z_0^{'2}(u)}=0~. 
	\end{align}
	The first differential equation is due to pure AdS spacetime and it has a solution of the form mentioned in eq.\eqref{z0(u)}. Then substituting the expression of $z_{0}(u)$ in the second differential equation, we can perform a similar series solution as discussed in the previous section. Hence the solution of $z_1(u)$ reads 
	\begin{equation}\label{z1 matter}
		z_1(u)=\C^2\paren{\frac{u^2}{2\C}-\frac{l^2}{2\C}-\frac{1}{20}\frac{u^4}{\C^2}-\frac{1}{4}\frac{l^2u^2}{\C^2}+\frac{3}{10}\frac{l^4}{\C^2}}~.
	\end{equation}
	Now to calculate the area functional, we have to substitute the values of $z_{0}(u)$ and $z_{1}(u)$ from eq.\eqref{z0(u)} and eq.\eqref{z1 matter} respectively. Then by expanding the terms of the integrand (eq.\eqref{Area mat}) up to $\mathcal{O}(\frac{u^4}{\C^2})$ and keeping terms up to $\mathcal{O}(\T{\rc})$, we can perform the integral, which gives
	\begin{align}
		A^{(mat)}=\Omega_{2}\left[\frac{l}{2\sqrt{\C}\z^2}+\frac{l}{2\C^{3/2}}+\T{\rc} \C^{3/2}\paren{\frac{l^5}{10\C^{5/2}}}\right]~.
	\end{align}
	Hence, from the RT formula the HEE of the matter-dominated universe is given by
	\begin{align}\label{HEE mat}
		S_{HEE}^{(mat)}=\frac{\Omega_2}{4G_4}\left[\frac{l}{2\sqrt{\C}\z^2}+\frac{l}{2\C^{3/2}}+\T{\rc} \C^{3/2}\paren{\frac{l^5}{10\C^{5/2}}}\right]~.
	\end{align}
	It is clear from the above equation that the first term appears due to the contribution of pure AdS geometry and the second term corresponds to correction to the HEE due to the presence of matter in the braneworld model. \\Now we will proceed to study the early and late time behaviour of the HEE in a matter-dominated universe. In this scenario, the Israel junction condition gives the radial position of the brane, which is given by
	\begin{equation}
		\z =\frac{z_i}{1+\tilde{\rc }^{1/3}(3/2)^{2/3}z_{i}\tau^{2/3}}
	\end{equation}
	where $z_{i}$ is the initial position of the brane. For the early-time scenario, we know that $z_{i}<<\T{\rc}^{1/3}$, hence the brane's position can be approximated by
	\begin{equation}\label{Brane_Position_matter}
		\z \approx z_{i}\paren{1-\paren{\frac{3}{2}}^{2/3}\T{\rc}^{1/3}\tau^{2/3}z_{i}}~.
	\end{equation}
	Now we will put this value of $\z$ in eq.\eqref{HEE mat}, this gives
	\begin{align}
		S_{HEE}^{(mat)}=& \frac{R^3 \Omega_{2}}{4G}\left[\frac{l^3}{3}\ci^{-3/2}\paren{1+\frac{3\zeta  z_i \tau^{2/3}}{2\ci}}+\frac{2}{5}l^5\ci^{-5/2}\paren{1+{\frac{5\zeta z_i\tau^{2/3}}{2\ci}}} \right.\nonumber\\&\left.+\T{\rc}\frac{l^5}{10\ci}\paren{1+\frac{\zeta z_i^2\tau^{2/3}}{\ci}}\right]
	\end{align}
	where $\zeta=2\paren{\frac{3}{2}}^{2/3}\T{\rc}^{1/3}z_i$. The above expression can now be written as
	\begin{equation}\label{S_mat_early}
		S_{HEE}^{(mat)}=\xi+\upsilon\tau^{2/3} 
	\end{equation} 
	where $\xi=\frac{l^3}{3\ci^{3/2}}+\frac{2l^5}{5\ci^{5/2}}+\T{\rc}\frac{l^5}{10\ci}$ and $\upsilon=\frac{l^3\zeta z_i^2}{2\ci^{5/2}}+\frac{l^5\zeta z_i^2}{\ci^{7/2}}+\T{\rc}\frac{l^5 \zeta z_i^2}{10\ci^2}$.\\
	Hence, in the early times of the matter-dominated era, HEE scales as $\tau^{2/3}$. This also suggests that in the early times of a matter-dominated universe, brane position and HEE have similar scaling behaviour with the cosmological time.\\
	\begin{figure}[ht!]
		\centering
		\includegraphics[width=0.7\linewidth]{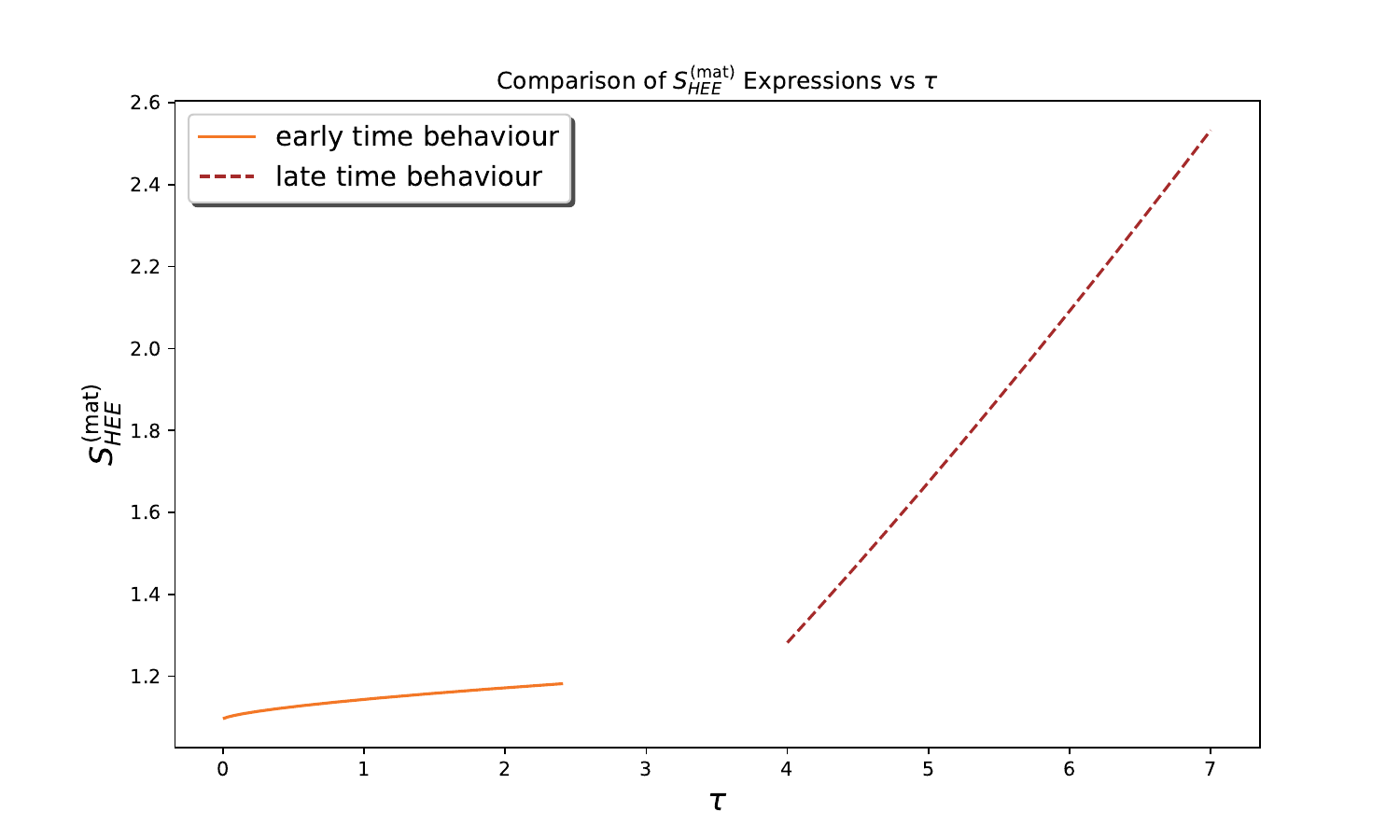}
		\caption{The above figure presents the dynamics of holographic entanglement entropy in the matter-dominated universe. Where the orange solid line represents the early time behaviour and the brown dashed line shows the late time growth.  While plotting we have taken $l=2$,$z_i=0.15$,$\T{\rc}=0.75$, keeping in mind that $z_i<<\T{\rc^-\frac{1}{3}}$. Here, the prefactor $\frac{\Omega_2}{4G}$ is set to 1.}
		\label{fig:S_mat}
	\end{figure}
	Now we will proceed to calculate the late time behaviour of the HEE. For late time ($\tau \to \infty$) of matter-dominated universe, the brane position is given by
	\begin{equation}\label{z bar mat late}
		\z(\tau) =\frac{(\frac{2}{3})^{2/3}}{\rc^{1/3}\tau^{2/3}}~. 
	\end{equation}
	Substituting this equation of $\z(\tau)$ in eq.\eqref{HEE mat}, we get the following expression of HEE for late time in matter-dominated universe
	\begin{equation}\label{S_mat_late}
		S^{(mat)}_{HEE}=\frac{\Omega_2}{4G}\left[\frac{\eta^{-2}\tau^{4/3}}{2}\left(1-\frac{\eta\tau^{-4/3}}{2l^2}\right)+\frac{1}{2l^2}\left(1-\frac{3\eta^2 \tau^{-4/3}}{2l^2}\right)+\frac{\T{\rc}l^3}{10}\left(1-\frac{\eta^2 \tau^{-4/3}}{l^2}\right)\right]
	\end{equation}
	where $\eta=\frac{(\frac{2}{3})^{2/3}}{\T{\rc}^{1/3}}$. Hence, in the late time era of the matter-dominated universe, the leading order term of HEE scales as $\tau^{4/3}$. The leading order behaviour of the HEE in the matter-dominated era also suggests that the HEE grows faster compared to the radiation-dominated universe.  In the late time of the matter-dominated universe, the physical length scale goes as $L\sim \tau^{2/3}$. Hence, following the area law, HEE should go as $L^2 \sim \tau^{4/3}$. Our result for $S^{(mat)}_{HEE}$ in the late time era, clearly suggests that it is consistent with the area law of HEE. To visualise things in a better way we have illustrated the early (eq.\eqref{S_mat_early}) and late time form (eq.\eqref{S_mat_late}) of the holographic entanglement entropy in a matter dominated universe in Fig.\eqref{fig:S_mat}. While obtaining the plot, we have chosen the parameters such that they obey $z_i<<\T{\rc^{-\frac{1}{3}}}$. One can clearly see that the solid orange line is growing like a $\tau^\frac{2}{3}$ and the dashed brown curve shows a $\tau^\frac{4}{3}$ behaviour for the very late time.
	\subsection{Holographic entanglement entropy in the presence of exotic matter}
	In this section, we will calculate the time dependence of the HEE for the braneworld model only having cosmic strings or some kind of exotic matter. In this case, the lapse function of the black brane will have a form $f(z(u)=1-\T{\delta}z(u)^2$, where $\T{\delta}=\frac{\delta}{R^2}$. Now reparameterizing the brane's radial position as a function of a coordinate on the brane and setting $dt=0$. The area functional corresponding to the black brane geometry (eq.\eqref{BB metric}) in the presence of exotic matter reads
	\begin{equation}\label{Area ex}
		A^{ex}=\Omega_{2}\int_0^ldu\frac{u^2}{z(u)^3}\sqrt{1+\frac{z'^2}{f_\kappa\paren{z(u}}}.
	\end{equation}
	In this case, the eular-Lagrange equation for $z(u)$ is given by
	\begin{align}\label{EL z exotic}
		& 3 u \left( 1 + (z'(u))^2 \right) + z(u) \left( 2 \left( z'(u) + (z'(u))^3 \right) + u z''(u) \right)\\&+3 \delta^2 u z(u)^4
		- 2 \delta u z(u)^2 \left( 3 + (z'(u))^2 \right) 
		- \delta z(u)^3 \left( 2 z'(u) + u z''(u) \right)=0~.
	\end{align}
	The above equation is quite difficult to solve analytically, hence just like the previous two matter sources we will choose a perturbative form for $z(u)$, which reads
	\begin{equation}
		z(u)=z_{0}(u)+\T{\delta}z_{1}(u)+\mathcal{O}(\T{\delta}^2)~.
	\end{equation}
	Now substituting the above expression of $z(u)$ in eq.\eqref{EL z exotic}, we get two differential equations corresponding to $\mathcal{O}(\T{\delta}^0)$ and $\mathcal{O}(\T{\delta}^1)$, which gives
	\begin{equation}
		\left( 3 u + 2 z_0(u) z_0'(u) \right) \left( 1 + (z_0'^2(u) \right) + 
		u z_0(u) z_0''(u)=0
	\end{equation}
	\begin{align}
		&
		u z_1(u) z_0''(u) 
		- z_0(u)^3 \left( 2 z_0'(u) + u z_0''(u) \right) 
		+ z_0(u) \left( (2 + 6 (z_0'(u))^2) z_1'(u) + u z_1''(u) \right)\nonumber\\&-2 u z_0(u)^2 \left( 3 + (z_0'(u))^2 \right) 
		+ 2 z_0'(u) \left( z_1(u) \left( 1 + (z_0'(u))^2 \right) + 3 u z_1'(u) \right) =0~.
	\end{align}
	As mentioned earlier, the solution of the first equation is given by
	\begin{equation}\label{ex z0}
		z_0(u)=\sqrt{\C-u^2}~.
	\end{equation}
	Substituting the solution of $z_{0}(u)$ in the second differential equation, we can again perform a series solution for $z_{1}(u)$. The solution can be written as 
	\begin{align}\label{ex z1}
		z_1(u)&=\delta(\C^{3/2}\left(\frac{u^2}{2\C}-\frac{l^2}{2\C}+\frac{1}{20}\frac{u^4}{\C^2}+\frac{l^4}{5\C^2}-\frac{1}{4}\frac{l^2u^2}{\C^2}\right.\nonumber \\&\left.+\mathcal{O}\paren{\frac{u^6}{\C^3}}+\mathcal{O}\paren{\frac{l^6}{\C^3}}+\mathcal{O}\paren{\frac{u^4l^2}{\C^3}}+\mathcal{O}\paren{\frac{u^2l^4}{\C^3}}\right)~.
	\end{align}
	Hence, the form of the $z(u)$  becomes
	\begin{align}\label{z ex}
		& z(u)=\sqrt{\C}\paren{1-\frac{1}{2}\frac{u^2}{\C}-\frac{1}{8}\frac{u^4}{\C^2}+\mathcal{O}(\frac{u^6}{\C^3})}\nonumber \\&+\delta(\C^{3/2}\left(\frac{u^2}{2\C}-\frac{l^2}{2\C}+\frac{1}{20}\frac{u^4}{\C^2}+\frac{l^4}{5\C^2}-\frac{1}{4}\frac{l^2u^2}{\C^2}\right)~.
	\end{align}
	In order to calculate the area functional, we will again put the values of $z_{0}(u)$ and $z_{1}(u)$ from eq(s).(\eqref{ex z0},\eqref{ex z1}) in eq.\eqref{Area ex}. Before calculating the integral, we will expand the integrand up to $\mathcal{O}(\frac{u^4}{\C^2})$, and then the final expression of the area of the RT surface reads
	\begin{equation}
		A^{ex}=\Omega_{2}\left[\frac{l}{2\sqrt{\C}\z^2}+\frac{l}{2\C^{3/2}}+\delta\C\paren{\frac{1}{10}\frac{l^5}{\C^{5/2}}+\frac{51}{70}\frac{l^7}{\C^{7/2}}}\right]~.
	\end{equation}
	Again using the RT proposal the HEE of the universe for an exotic matter-dominated universe can be written as
	\begin{equation}\label{HEE ex}
		S^{ex}_{HEE}=\frac{\Omega_{2}}{4G}\left[\frac{l}{2\sqrt{\C}\z^2}+\frac{l}{2\C^{3/2}}+\delta\C\paren{\frac{1}{10}\frac{l^5}{\C^{5/2}}+\frac{51}{70}\frac{l^7}{\C^{7/2}}}\right]~.
	\end{equation}
	In the early time of the exotic matter-dominated era, the brane position can be obtained from eq.\eqref{bar r ex} , which is given by
	\begin{equation}\label{z bar ex early}
		\z(\tau)=z_{i}\paren{1-\sqrt{\T{\delta}}z_{i}\tau}~.
	\end{equation}
	In this early time regime, the HEE can be calculated by putting the above-mentioned form of $\z(\tau)$ in eq.\eqref{HEE ex}, which reads
	\begin{align}\label{S_ex early time}
		S^{ex}_{HEE}&=\frac{\Omega_{2}}{4G}\Bigg[\frac{l}{2z_{i}^2 \sqrt{\ci}}(1+\eta_{1}\tau)(1+2\sqrt{\T{\delta}}z_{i}\tau)+\frac{l}{2\ci^{3/2}}(1+3\eta_{1}\tau)\nonumber\\
		&\quad\Bigg.+\T{\delta}\paren{\frac{l^5}{10\ci^{3/2}}(1+3\eta_{1}\tau)+\frac{51 l^7}{70 \ci^{5/2}}(1+5\eta_1 \tau)}\Bigg]
	\end{align}
	where $\eta_{1}=\frac{\sqrt{\T{\delta}}z_{i}^3}{2\ci}$.\\
	From the above expression of HEE, it is clear that in the early time of the exotic matter-dominated era, the leading order behaviour of HEE changes as $\tau$. Again, we can see that in the early time of an exotic matter-dominated universe, the leading order term of HEE and brane position scales in a similar fashion.\\
	\begin{figure}[h!]
		\centering
		\includegraphics[width=0.7\linewidth]{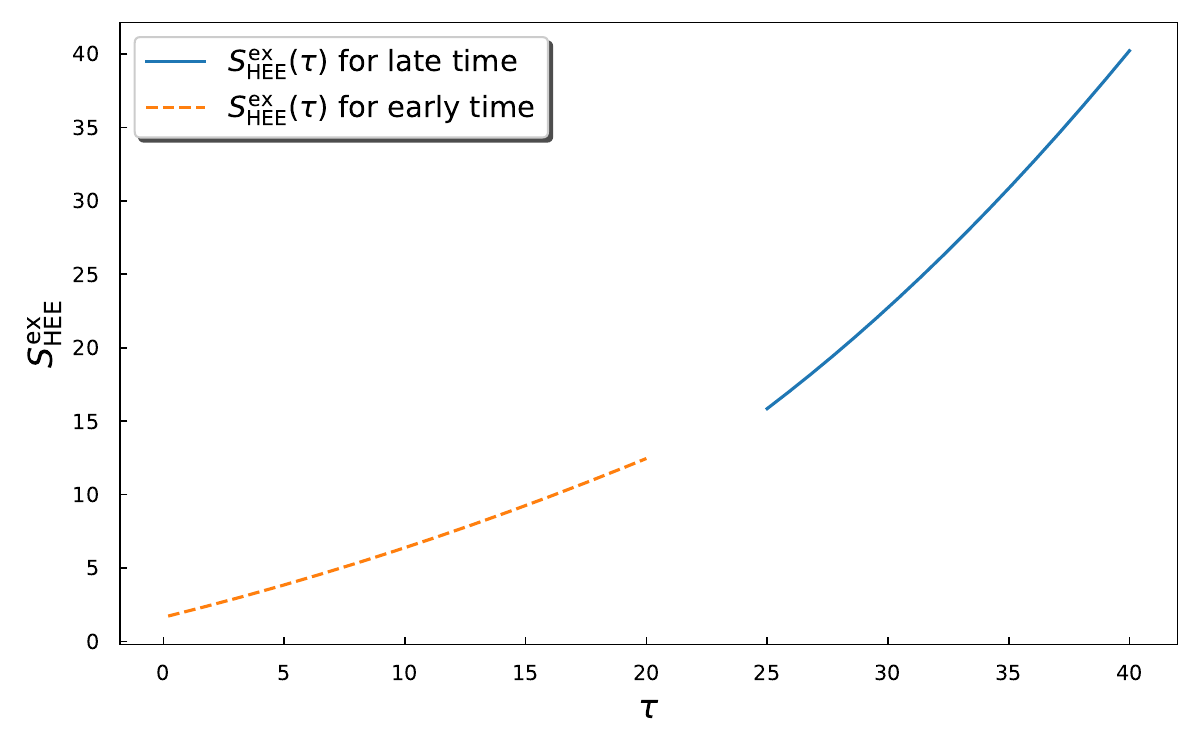}
		\caption{The above plot shows the variation of holographic entanglement entropy with cosmological time for early and late times of some exotic matter-dominated universe. The plots are done for $l=2$, $\T{\delta}=0.05$ and $z_i =1$. The dashed line in orange represents the early time behaviour, and the solid line depicts the late time behaviour of some exotic matter-dominated universe. Here, the prefactor $\frac{\Omega_2}{4G}$ is set to 1.}
		\label{fig:S_ex}
	\end{figure}
	Now we will proceed to compute the late time behaviour of the HEE in this case.
	In late time era, when $\tau\to\infty$, the brane position becomes
	\begin{equation}\label{z bar ex late}
		\z(\tau)=\frac{1}{\sqrt{\T{\delta}}\tau}~.
	\end{equation}
	Considering the above expression of $\z(\tau)$ in eq.\eqref{HEE ex}, we obtain
	\begin{align}\label{S_ex late time}
		S^{ex}_{HEE}&=\frac{\Omega_2}{4G}\Bigg[\frac{\T{\delta}\tau^2}{2}\paren{1-\frac{1}{2\T{\delta}l^2 \tau^2}}+\frac{1}{2l^2}\paren{1-\frac{3}{2\T{\delta}l^2 \tau^2}}\nonumber\\
		&\quad\Bigg.+\T{\delta}\paren{\frac{l^2}{10}\paren{1-\frac{3}{2\T{\delta}l^2 \tau^2}}+\frac{51 l^2}{70}\paren{1-\frac{5}{2\T{\delta}l^2 \tau^2}}}\Bigg]
	\end{align}
	Hence, the leading order of HEE in the late time of the exotic matter-dominated era grows as $\tau^2$.  This implies that for the late time of a universe dominated by exotic matter, HEE grows faster compared to radiation and matter-dominated universes. We would also like to mention that in the late time of an exotic matter-dominated universe, the physical length scale grows as $L\sim \tau$. This suggests HEE will change as $L^2 \sim \tau^2$. The leading order behaviour of $S^{ex}_{HEE}$ in the above equation confirms the validity of the area law in our result.\\ 
	We have also graphically represented our results (eq.(s)(\eqref{S_ex early time},\eqref{S_ex late time})) in Fig.\eqref{fig:S_ex}. In obtaining Fig.\eqref{fig:S_ex}, we have chosen $l=2$, $\T{\delta}=0.05$ and $z_i =12$. We have chosen these parameters such that they obey the relation In the Figure, the orange dashed line represents the early time dynamics, and the solid blue line shows the late time dynamics of HEE for an exotic matter-dominated universe. From the graph, it can be clearly seen that after a certain time, the late time expression of HEE started dominating the early time behaviour.
	
	\section{Complexity of the universe}\label{Complexity of the universe}
	In this section, we will calculate the complexity of the universe for different matter sources and show their early and late time dependence. The calculations are done for a spherical small subsystem on the brane.
	\subsection{Time independent volume complexity for pure AdS spacetime}
	We will start this subsection with a discussion of
	various holographic proposals aimed to calculate the quantum complexity of boundary states. In \cite{Susskind:2014moa,Susskind:2014rva,Susskind:2018pmk}, it was suggested that the complexity of the boundary states is directly related to the volume of the Einstein-Rosen bridge (ERB) that links the boundaries of a two-sided eternal black hole. According to this proposal, the complexity is given by
	\begin{equation}
		C_V (t_{L},t_{R})=\frac{V^{ERB}(t_L ,t_R)}{8\pi R G_{d+1}}
	\end{equation}
	where R is the AdS radius and $V^{ERB}(t_L,t_R)$ is the co-dimension one extremal volume of ERB which is bounded by the two spatial slices at times $t_L$ and $t_R$ of two CFTs that live on the boundaries of the eternal black hole. This equation is referred to as the "Complexity=Volume" conjecture.\\
	Another proposal \cite{Alishahiha:2018lfv,Brown:2015bva,Brown:2015lvg,Goto:2018iay} suggests that complexity can be derived from the bulk action assessed on the Wheeler-DeWitt patch, which is enclosed by light sheets. This is referred to as the "Complexity = Action" conjecture. This conjecture gives the following relation of complexity
	\begin{equation}
		C_{A}=\frac{I_{WdW}}{\pi \hbar}~.
	\end{equation}
	In our analysis, we will use the "Complexity=Volume" conjecture \cite{Stanford:2014jda,Alishahiha:2015rta}. This conjecture tells us that holographic subregion complexity is proportional to the volume enclosed by the minimal RT surface in the bulk, then the expression for holographic subregion complexity becomes
	\begin{equation}
		C_{V}=\frac{V_{\gamma}}{8\pi RG_{d+1}}
	\end{equation}
	where $V_{\gamma}$ is the volume enclosed by minimal hypersurface in the bulk. \\We will now proceed to compute the time-independent HSC of the FLRW universe in the absence of any matter sources. As discussed earlier the bulk geometry should be chosen as pure AdS spacetime. Hence, for the spacetime metric in eq.\eqref{AdS metric}, the volume enclosed by the RT surface and the brane is given by
	\begin{equation}
		V^{(AdS)}=\Omega_{2}\int^{l}_{0}du u^{2}\int_{z(u)}^{\z}\frac{dz}{z^{4}}=\Omega_{2}\int_{0}^{l}du u^2\paren{\frac{1}{3z(u)^3}-\frac{1}{3\z^3}}~.
	\end{equation}
	After putting the expression of $z(u)$ in the above equation and doing a series expansion of $z(u)$ in the integrand, we can easily perform the integral, which gives
	\begin{equation}
		V^{(AdS)}=\Omega_{2}\left[\frac{l^3}{9\C^{3/2}}+\frac{l^5}{10\C^{5/2}}+\frac{5l^7}{56\C^{7/2}}-\frac{l^3}{9\z^3}+\mathcal{O}\paren{\frac{l^9}{\C^{9/2}}}\right]~.
	\end{equation}
	Now according to the "Complexity=Volume" conjecture \cite{Alishahiha:2015rta}, the complexity of the universe is given by
	\begin{equation}\label{C AdS}
		C^{(AdS)}_{V}=\frac{\Omega_{2}}{8\pi G}\left[\frac{l^3}{9\C^{3/2}}+\frac{l^5}{10\C^{5/2}}+\frac{5l^7}{56\C^{7/2}}-\frac{l^3}{9\z^3}+\mathcal{O}\paren{\frac{l^9}{\C^{9/2}}}\right]~.
	\end{equation}
	The above expression denotes the form of HSC for a static universe in absence of any kind of matter source. Now we will proceed to compute to HSC for an eternally inflating universe.\\
	In case of an eternally inflating universe, the brane's radial position is given by eq(\ref{z bar eter early})
	\begin{align}
		C^{(AdS)}_{V}&=\frac{\Omega_{2}}{8\pi G}\Bigg[\frac{l^3 r_{0}^3}{9(l^2 r_{0}^2 +1)^{3/2}}\paren{1+\frac{3 H\tau}{l^2 r_{0}^2 +1}}+\frac{l^5 r_{0}^5}{10(l^2 r_{0}^2 +1)^{5/2}}\paren{1+\frac{5 H\tau}{l^2 r_{0}^2 +1}}\nonumber\\
		&\quad\Bigg.+\frac{5l^7 r_{0}^7}{56(l^2 r_{0}^2 +1)^{3/2}}\paren{1+\frac{3 H\tau}{l^2 r_{0}^2 +1}}-\frac{l^3 r_{0}^3}{9}(1+3H\tau)\Bigg]~.
	\end{align}
	Hence, the volume complexity of an eternally inflating universe in early time changes as $C^{(AdS)}_{V}\sim\tau$. This result suggests that in the early time of an eternally inflating universe the leading order term of complexity and radial position of the brane have similar scaling behavior with cosmological time ($\tau$).\\
	Now we will proceed to see how the complexity of an eternally inflating universe changes with cosmological time in the late time era. 
	In the late time era of an eternally inflating universe, the brane position is approximately given by eq.\eqref{z bar eter late}. Therefore, using this expression of $\z(\tau)$ in eq.\eqref{C AdS}, we obtain
	\begin{align}
		C^{(AdS)}_{V}&=\frac{\Omega_{2}}{8\pi G}\Bigg[\frac{1}{9}\paren{1-\frac{3e^{-2H\tau}}{2l^2 r_{0}^2}}+\frac{1}{10}\paren{1-\frac{5e^{-2H\tau}}{2l^2 r_{0}^2}}+\frac{5}{56}\paren{1-\frac{7e^{-2H\tau}}{2l^2 r_{0}^2}}-\frac{l^3 r_{0}^3}{9}e^{3H\tau}\Bigg]
	\end{align}
	The leading order of the above equation changes with cosmological time as $e^{3H\tau}$. Hence, in the late time era of the eternally inflating universe, the volume complexity changes as $C^{(AdS)}_{V}\sim e^{3H\tau}$. As the physical length scale of the universe for an eternally inflating universe changes as $L\sim e^{H\tau}$, the volume law suggests that complexity should scale as $L^3 \sim e^{3H\tau}$. From the above expression of $C^{(AdS)}_{V}$, in the late time, it is clear that our result agrees with the volume law of holographic complexity.
	\subsection{Holographic complexity of universe for radiation dominated era}
	For a radiation-dominated universe, the blackening factor of the black brane geometry (eq.\eqref{BB metric}) is given by $f(z(u))=1-\m z(u)^4$. Hence in the presence of radiation in the braneworld model, the volume enclosed by the RT surface is given by
	\begin{equation}
		V^{(rad)}=\Omega_2\int^{l}_{0}du u^{2}\int_{z(u)}^{\z}\frac{dz}{z^{4}\sqrt{f(z)}}
	\end{equation}
	where $f(z(u))=1-\m z(u)^4$. Here, $z(u)=z_{0}(u)+\m z_{1}(u)$, thus to make our calculation simpler we will put this expression of $z(u)$ and expand the integrand up to $\mathcal{O}(\m)$. Then in order to do the integral, we will substitute the values of $z_{0}(u)$ and $z_{1}(u)$ from eq.(s)(\eqref{z0(u)},\eqref{z1 eq rad}). We will then expand the integrand, retaining terms up to $\mathcal{O}(\frac{u^4}{\C^2})$. This leads us to the following expression for the volume enclosed by the RT surface and the brane
	\begin{align}
		V^{(rad)}& =\Omega_{2}\left[\left(\frac{l^{3}}{9\C^{3/2}}+\frac{l^5}{10 \C^{5/2}}+\frac{5 l^7}{56 \C^{7/2}}-\frac{g l^3}{3}\right)\right.\\&\left. +\m \C^{4}\left(-\frac{l^3}{6 \C^{3/2}}+\frac{7 l^5}{60 \C^{5/2}}-\frac{7 l^7}{240 \C^{7/2}}\right)\right]~.
	\end{align}
	Now as we have the expression of $V^{(rad)}$, we can easily compute the complexity from the "complexity=volume" conjecture \cite{Alishahiha:2015rta}.
	The volume complexity is given by 
	\begin{align}
		C^{(rad)}_{V}&=\frac{V_{Rad}}{8\pi G_{4}}\nonumber\\
		& =\frac{R^{4}\Omega_{2}}{8\pi G_{4}}\left[\left(\frac{l^{3}}{9\C^{3/2}}+\frac{l^5}{10 \C^{5/2}}+\frac{5 l^7}{56 \C^{7/2}}-\frac{g l^3}{3}\right)\right.\\&\left. +\m \C^{4}\left(-\frac{l^3}{6 \C^{3/2}}+\frac{7 l^5}{60 \C^{5/2}}-\frac{7 l^7}{240 \C^{7/2}}\right)\right]
	\end{align}
	where $g=-\frac{1}{3\z^3}+\frac{\m}{2}\z$. Now to observe the time dependence of the complexity in the early and late time eras, we must use the expression for the brane position given in eq.\eqref{rad z bar}.\\
	In the late time regime($\tau\rightarrow \infty$), we can approximately write the expression of the brane position ($\z$) as 
	\begin{equation}
		\z \simeq \frac{z_h}{\sqrt{2\tau}}\paren{1-\frac{z_h}{z_i\sqrt{2\tau}}}~.
	\end{equation} 
	Inserting this into the expression of complexity, one gets 
	\begin{equation}
		V^{(rad)}=V_{0}+V_{1}
	\end{equation}
	where $V_0$ and $V_1$ is given by
	\begin{align}
		& V_0=\Omega_{2}\paren{-\frac{439}{840}-\frac{1}{\tau}\paren{\frac{17z_h^2}{32l^2}}+\frac{l^32^{3/2}}{9z_h^3}\tau^{3/2}+\frac{2l^3}{3z_h^2z_i}\tau}
	\end{align}
	\begin{align}
		V_1 = \T{m}l^4\paren{-\frac{19}{240}-\frac{19z_h^2}{240 l^2}\frac{1}{\tau}+\frac{29}{240}\frac{1}{\tau}}+\T{m}\frac{l^3}{6}\paren{-\frac{z_h}{\sqrt{2\tau}}+\frac{z_h^2}{2\tau z_i}}
	\end{align}
		\begin{figure}[h!]
		\centering
		\includegraphics[width=0.7\linewidth]{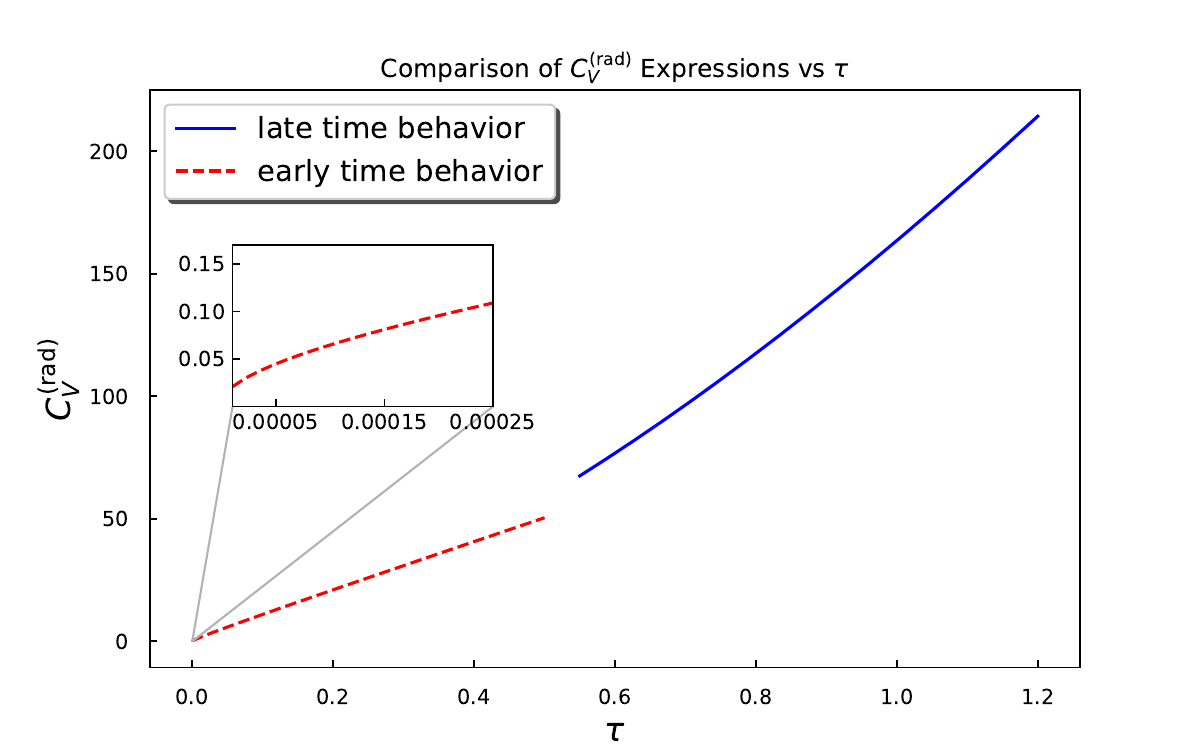}
		\caption{The above plot shows the variation of volume complexity with respect to the cosmological time for the early and late time regimes of the radiation-dominated universe. This plot is done for the parameters $l=2$, $z_h =0.2$, $z_i =10$ and $\T{m}=0.01$. The dashed plot in red indicates the early-time behavior of holographic complexity of the radiation-dominated era in the early time. The solid line in blue denotes the late-time behavior of the complexity in the radiation-dominated era. Here, the prefactor $\frac{\Omega_2}{4G}$ is set to 1.}
		\label{fig:cv rad}
	\end{figure}
	Hence, the form of HSC for the radiation-dominated era in early times becomes
	\begin{align}
		C^{(rad)}_{V}=&\frac{\Omega_2}{8\pi G}\left[-\frac{439}{840}-\frac{1}{\tau}\paren{\frac{17z_h^2}{32l^2}}+\frac{l^32^{3/2}}{9z_h^3}\tau^{3/2}+\frac{2l^3}{3z_h^2z_i}\tau\right.\nonumber\\&\left.+\T{m}l^4\paren{-\frac{19}{240}-\frac{19z_h^2}{240 l^2}\frac{1}{\tau}+\frac{29}{240}\frac{1}{\tau}}+\T{m}\frac{l^3}{6}\paren{-\frac{z_h}{\sqrt{2\tau}}+\frac{z_h^2}{2\tau z_i}}\right]~.
	\end{align}
	From the above expression, it is clear that in late time the HSC of the universe in the radiation-dominated era is proportional to $\tau^{3/2}$. In a radiation-dominated universe, the late-time behaviour of the length scale is $L\sim \tau^{1/2}$. Hence, the volume complexity should be proportional to $\tau^{3/2}$. Thus, our computed expression of complexity is consistent with the volume law. 
	Hence, it is evident that in radiation dominated universe complexity grows faster compared to HEE.\\
	In the early time of the radiation-dominated universe, the position of the brane is given by eq.\eqref{z bar rad early}. Therefore the total volume enclosed by the RT surface is given by the following expression
	The volume enclosed by the RT surface is given by
	\begin{equation}
		V=V_{0}+V_{1}
	\end{equation}
	where the expression of $V_0$ is given by
	\begin{align}
		& V_0\sim \paren{\frac{l^3}{9\ci^{3/2}}+\frac{l^5}{10\ci^{5/2}}+\frac{5l^7}{56\ci^{7/2}}}\nonumber\\&+\frac{\sqrt{2}z_i^3}{z_h\ci}\tau\paren{\frac{l^3}{\ci^{3/2}}+\frac{l^5}{2\ci^{5/2}}+\frac{5}{8}\frac{l^7}{\ci^{7/2}}}+\frac{l^3}{9z_i^2}\paren{1+\frac{2\sqrt{2}z_i}{z_h}\sqrt{\tau}}
	\end{align}
	and the expression of $V_1$ is given by
	\begin{align}
		& V_1\sim\T{m}\ci^2\left(-\frac{l^3}{3\ci^3}\paren{1-\frac{3\sqrt{2}z_i^3}{z_h\ci}\tau}+\frac{7}{60}\frac{l^5}{\ci^{5/2}}\paren{1-5\sqrt{2}\frac{z_i^3}{z_h\ci}\tau}\right.\nonumber\\&\left.-\frac{7}{240}\frac{l^7}{\ci^{7/2}}\paren{1-7\sqrt{2}\frac{z_i^3}{z_h\ci}\sqrt{\tau}}\right)-\frac{m}{2}z_i\paren{1-\frac{\sqrt{2}z_i}{z_h}\sqrt{\tau}}\nonumber\\& +\T{m}\frac{4\sqrt{2}z_i\ci}{z_h}\tau\paren{\frac{l^3}{3\ci^{3/2}}-\frac{7}{60}\frac{l^5}{\ci^{5/2}}+\frac{7}{240}\frac{l^7}{\ci^{7/2}}}~.
	\end{align}
	Hence, the final expression of HSC becomes
	\begin{align}
		&C^{\text{(rad)}}_{V} =  \frac{\Omega_{2}}{8\pi G} \left[ \left( \frac{l^3}{9\ci^{3/2}} + \frac{l^5}{10\ci^{5/2}} + \frac{5l^7}{56\ci^{7/2}} \right) \right. \nonumber \\
		& \left. + \frac{\sqrt{2}z_i^3}{z_h \ci} \tau \left( \frac{l^3}{\ci^{3/2}} + \frac{l^5}{2\ci^{5/2}} + \frac{5}{8} \frac{l^7}{\ci^{7/2}} \right) \right. \nonumber \\
		& \left. + \frac{l^3}{9z_i^2} \left( 1 + \frac{2\sqrt{2}z_i}{z_h} \sqrt{\tau} \right) - \T{m} \ci^2 \left( \frac{l^3}{3 \ci^3} \left( 1 - \frac{3\sqrt{2}z_i^3}{z_h \ci} \tau \right) \right. \right. \nonumber \\
		& \left. \left. - \frac{7}{60} \frac{l^5}{\ci^{5/2}} \left( 1 - 5\sqrt{2} \frac{z_i^3}{z_h \ci} \tau \right) + \frac{7}{240} \frac{l^7}{\ci^{7/2}} \left( 1 - 7\sqrt{2} \frac{z_i^3}{z_h \ci} \sqrt{\tau} \right) \right) \right. \nonumber \\
		& \left. + \frac{\T{m}}{2} z_i \left( 1 - \frac{\sqrt{2} z_i}{z_h} \sqrt{\tau} \right) - \T{m} \frac{4 \sqrt{2} z_i \ci}{z_h} \tau \left( \frac{l^3}{3 \ci^{3/2}} \right.\right.\\&\left.\left.- \frac{7}{60} \frac{l^5}{\ci^{5/2}} + \frac{7}{240} \frac{l^7}{\ci^{7/2}} \right) \right] ~.
	\end{align}
	It can be seen that for the early time of the radiation-dominated universe, both complexity and radial position of the brane change in a similar fashion with the cosmological time. Now, having the expressions of early and late time complexity of radiation-dominated universe in hand, we have also shown our results graphically in Fig.\eqref{fig:cv rad}. In obtaining the graph, we have taken the parameters $l=2$, $z_h =0.2$, $z_i =10$ and $\T{m}=0.01$.  
	The evolution of the volume complexity has been shown in Figure \eqref{fig:cv rad})for early and late time regimes. For the early time (that is, the red dashed line), one can see that for very early times it grows as $\sqrt{\tau}$, and to visually illustrate it more clearly, we have zoomed it into an inset plot.  It is clear from the above expression that in the early time of the radiation-dominated universe, the leading order term of HSC scales as $\sqrt{\tau}$. 
	\subsection{Holographic complexity of universe for matter-dominated era}
	In this section, we will calculate the HSC for the matter-dominated universe. To do so we will first calculate the volume enclosed by the minimal RT surface. Following the black brane geometry in eq.\eqref{BB metric}, the volume enclosed by the static minimal RT surface is given by
	\begin{equation}
		V^{(rad)}=\Omega_2 \int^{l}_{0}du u^{2}\int_{z(u)}^{\z}\frac{dz}{z^{4}\sqrt{f(z)}}
	\end{equation}
	where $f(z(u))=1-\T{\rc}z^3$.

	\vspace{0.5cm}
	\noindent Now by substituting the expressions of $z_{0}(u)$ and $z_{1}(u)$ from eq(s).(\eqref{z0(u)},\eqref{z1 matter}) in the above equation and expanding the integrand up to $\mathcal{O}(\frac{u^4}{\C^2})$, we can easily evaluate the above integral. This gives the following expression of the volume
	\begin{align}
		V^{(mat)}&=\Omega_2\left[\left(\frac{l^3}{9\C^{\frac{3}{2}}}+\frac{l^5}{10 \C^\frac{5}{2}}+\frac{5l^7}{56\C^\frac{7}{2}}-\frac{gl^3}{3}\right) \nonumber \right.\\&\left. + \rc \C^{3/2} \left(\frac{7}{60}\frac{l^5}{\C^{5/2}}+\frac{9}{280}\frac{l^7}{\C^{7/2}}-\frac{1}{6}\frac{l^3}{\C^{3/2}}ln\left(\sqrt{\C}\right)\right)\right]
	\end{align}
	where $g=\frac{1}{3\z^3}-\frac{\T{\rc}}{2}log \z$.
	
	\vspace{0.4cm}
	\noindent Now, from the "Complexity=Volume" conjecture, we can say that the HSC of the matter-dominated universe is given by
	\begin{align}\label{C mat}
		C^{(mat)}_{V}&=\frac{\Omega_2}{8\pi G}\left[\left(\frac{l^3}{9\C^{\frac{3}{2}}}+\frac{l^5}{10 \C^\frac{5}{2}}+\frac{5l^7}{56\C^\frac{7}{2}}-\frac{gl^3}{3}\right) \nonumber \right.\\&\left. + \T{\rc} \C^{3/2} \left(\frac{7}{60}\frac{l^5}{\C^{5/2}}+\frac{9}{280}\frac{l^7}{\C^{7/2}}-\frac{1}{6}\frac{l^3}{\C^{3/2}}ln\left(\sqrt{\C}\right)\right)\right]~.
	\end{align}
	With this expression of $C^{(mat)}$, we will now observe the early and late time behaviour of the HSC for matter-dominated universe. For early time we will use the value of the brane's position form eq.(\ref{Brane_Position_matter}) and substitute it in the above equation. This gives
	\begin{align}\label{C_V_Mat_Early}
		&C^{(mat)}_{V}=\frac{\Omega_2}{8\pi G}\Bigg[\frac{l^3}{9\ci^{3/2}}\paren{1+\frac{3\xi \tau^{2/3}}{\ci}}+\frac{l^5}{10\ci^{5/2}}\paren{1+\frac{5\xi \tau^{2/3}}{\ci}}\nonumber\\
		&\quad\Bigg.+\frac{5 l^7}{56\ci^{7/2}}\paren{1+\frac{7\xi \tau^{2/3}}{\ci}}+\frac{l^3}{3}\paren{-\frac{1}{3z_{i}^2}\paren{1+\frac{3\xi\tau^{2/3}}{\ci}}+\frac{\T{\rc}}{2}log\paren{z_{i}(1-\frac{\xi\tau^{2/3}}{z_{i}^2})}}\nonumber\\
		&\quad\Bigg.+\T{\rc}\ci^{3/2}\paren{1-\frac{3\xi\tau^{2/3}}{\ci}}\left(\frac{7l^5}{60\ci^{5/2}}\paren{1+\frac{5\xi\tau^{2/3}}{\ci}}\nonumber\right.\\&\left.+\frac{9l^7}{280\ci^{7/2}}\paren{1+\frac{7\xi\tau^{2/3}}{\ci}}-\frac{l^3 \paren{1+\frac{3\xi\tau^{2/3}}{\ci}}}{6\ci^{3/2}}log\paren{\sqrt{\ci}\paren{1-\frac{\xi\tau^{2/3}}{\ci}}}\right)\Bigg]~
	\end{align}
	where $\xi=(3/2)^{2/3}\rc^{1/3}z_i^3$. From the above equation, we can clearly see that the leading order of $C^{(mat)}_{V}$ in the early time of the matter-dominated universe changes as $\tau^{2/3}$. Thus the radial velocity of the brane with respect to cosmological time and the growth rate of complexity show similar behaviour in the leading order in cosmological time.\\ 
	\begin{figure}[ht!]
		\centering
		\includegraphics[width=0.7\linewidth]{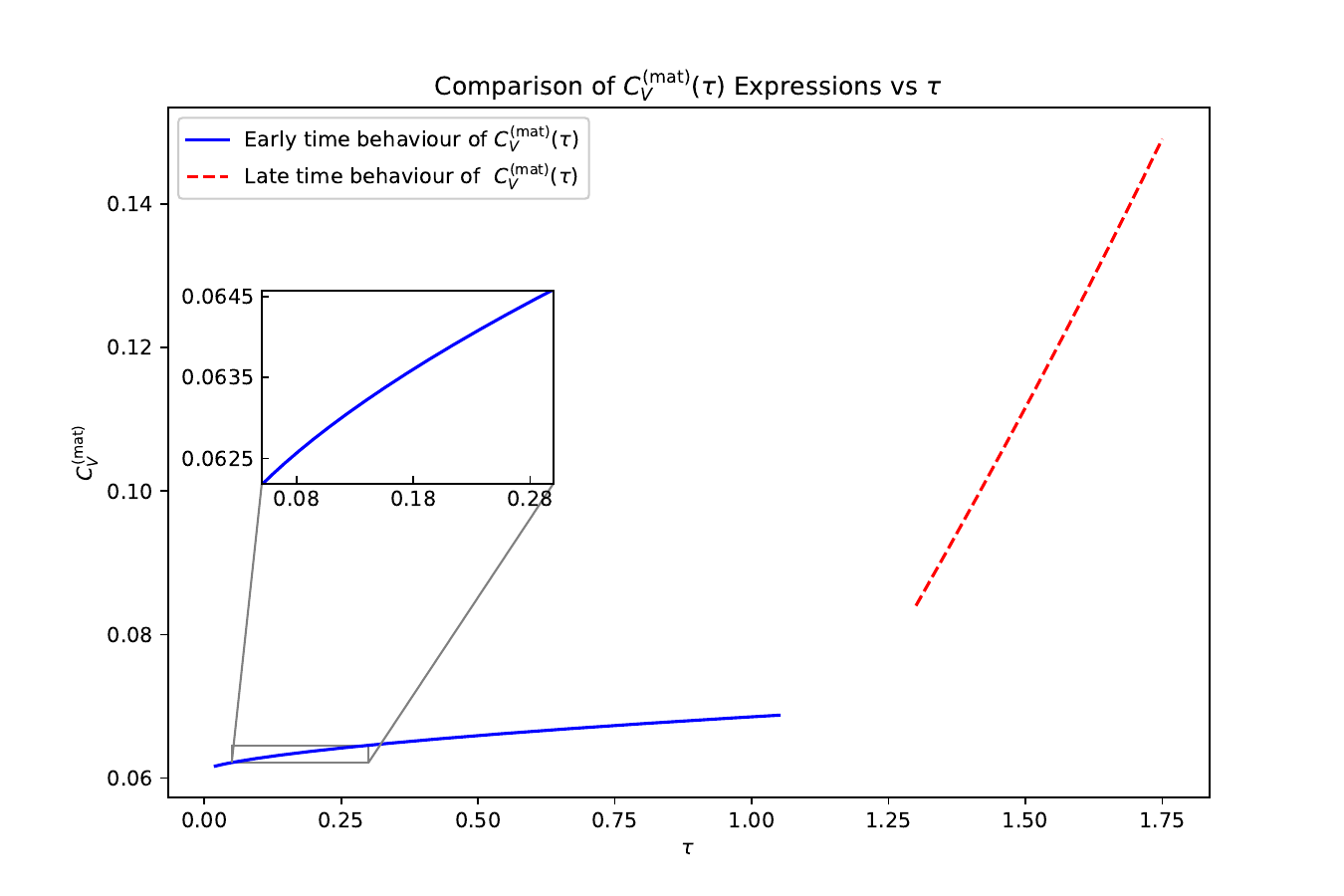}
		\caption{The above plot presents the evolution of the volume complexity with respect to the cosmological time for a matter-dominated universe for early as well as late times. While plotting, we have considered the values of $l=2,z_i=2,\xi=0.1,\T{\rc}=0.5$. The blue solid line depicts the early time behaviour of the complexity, while the red dashed line corresponds to the late time dynamics of the complexity. Here, the prefactor $\frac{\Omega_2}{4G}$ is set to 1.}
		\label{fig:C_V_Mat}
	\end{figure}
	Now we will again proceed to calculate the late time behaviour of complexity in a matter-dominated universe. As we already know that the late time behaviour of the brane's position in a matter-dominated universe, we can substitute the expression of $\z(\tau)$ from eq.\eqref{z bar mat late} in eq.\eqref{C mat}. This gives
	\begin{align}\label{C_V_Mat_Late}
		C^{(mat)}_{V}&=\frac{\Omega_2}{8\pi G}\Bigg[\frac{1}{3}\paren{1-\frac{3\eta^2 \tau^{-4/3}}{2l^2}}+\frac{1}{10}\paren{1-\frac{5\eta^2 \tau^{-4/3}}{2l^2}}+\frac{5}{56}\paren{1-\frac{7\eta^2 \tau^{-4/3}}{2l^2}}\nonumber\\
		&\quad\Bigg.+\T{\rc}l^3 \paren{1+\frac{3\eta^2 \tau^{-4/3}}{2l^2}}\left(\frac{7}{60}\paren{1-\frac{5\eta^2 \tau^{-4/3}}{2l^2}}+\frac{9}{280}\paren{1-\frac{7\eta^2 \tau^{-4/3}}{2l^2}}\nonumber\right.\\&\left.-\frac{1}{6}\paren{1-\frac{3\eta^2 \tau^{-4/3}}{2l^2}}log\paren{l\paren{1+\frac{\eta^2 \tau^{-4/3}}{2l^2}}}\right)-\frac{l^3}{3}\paren{\frac{\tau^2}{3\eta^3}-\frac{\T{\rc}}{2}log(\eta\tau^{-2/3})}\Bigg]
	\end{align}
	where $\eta=\frac{\paren{\frac{2}{3}}^{2/3}}{\T{\rc}^{1/3}}$. \\
	Hence, we can say that the leading order of the volume complexity for late time of the matter-dominated era scales as $\tau^2$. Again, we know that the physical length scale for the late time of a matter-dominated universe scales as $L\sim \tau^{2/3}$, hence, according to the volume law, complexity should change as $L^3$. The leading term of our expression of $C^{(mat)}_{V}$ in the late time clearly indicates the validity of the volume law in our result. It should also be mentioned that the growth rate of complexity in this case is larger than in the case of a radiation-dominated universe.  The form of volume complexity for the early time era from eq.\eqref{C_V_Mat_Early} and the late time era from eq.\eqref{C_V_Mat_Late} have been plotted  Fig.\eqref{fig:C_V_Mat}, where the solid blue line corresponds to early time behaviour and the red dashed line indicates the late time dynamics. In the very early time, one can see from the inset plot that the volume complexity scales proportional to  $\tau^\frac{2}{3}$, and from the red dashed line, one can clearly see that for very late time it scales as $\tau^2$.
	\subsection{Holographic complexity in presence of exotic matter}\label{exotic matter}
	In this section, we will compute the volume complexity of the braneworld model in the presence of only some exotic matter. In the braneworld model, this exotic matter is realised by the back reaction of two-branes in the bulk. In this case, the lapse function of the black brane geometry is given by $f(z(u))=1-\T{\delta}z(u)^2$. Similar to the previous cases, the volume enclosed by the RT surface will be 
	\begin{equation}
		V^{ex}=\Omega_2 \int^{l}_{0}du u^{2}\int_{z(u)}^{\z}\frac{dz}{z^{4}\sqrt{f(z)}}
	\end{equation}
	where $f(z(u))=1-\T{\delta}z(u)^2$. Now we will put the expression of $z(u)$ from eq.\eqref{z ex}, then we will expand the integrand up to $\mathcal{O}(\frac{u^4}{\C^2})$. This will eventually help us to perform the integral easily. This leads us to the following expression of volume, enclosed by the RT surface
	\begin{equation}
		\begin{split}
			V^{ex}&=\Omega_{2}\left[ \frac{l^3}{9 \C^{3/2}} - \frac{g l^3}{3} + \frac{l^5}{10 \C^{5/2}} + \frac{5 l^7}{56 \C^{7/2}} \right. \\&
			\left.+ \delta \C^{3/2}\left( \frac{7 l^5}{60 \C^{5/2}} + \frac{9 l^7}{280 \C^{7/2}} - \frac{1}{12} l^3 \log \C \right)
			\right]~
		\end{split}
	\end{equation}
	where $g=-\frac{1}{3\z^3}+\frac{\T{\delta}}{2\z}$.
	Again, from the "Complexity=Volume" conjecture, we can write that the complexity of the exotic-dominated universe is given by
	\begin{equation}\label{C ex}
		\begin{split}
			C_V^{ex}=\frac{\Omega_2}{8\pi G}&\left[ \frac{l^3}{9 \C^{3/2}} - \frac{g l^3}{3} + \frac{l^5}{10 \C^{5/2}} + \frac{5 l^7}{56 \C^{7/2}} \right. \\&
			\left.+ \T{\delta} \C^{3/2}\left( \frac{7 l^5}{60 \C^{5/2}} + \frac{9 l^7}{280 \C^{7/2}} - \frac{1}{12} l^3 \log \C \right)
			\right]~.
		\end{split}
	\end{equation}
	With this expression of complexity in hand, we will now show the early and late time behaviour of complexity with cosmological time for an exotic matter-dominated era.\\
	In this scenario the early time behaviour of the brane position is governed by eq.\eqref{z bar ex early}. Substituting this form of $\z(\tau)$ in the above expression of complexity, we get
	\begin{align}\label{CV_ex_early}
		&C^{ex}_{V} = \frac{\Omega_2}{8\pi G} \Bigg[ \frac{l^3}{3\ci^{3/2}}\left(1+\frac{3\alpha\tau}{2}\right)+\frac{l^5}{10\ci^{5/2}}\left(1+\frac{5\alpha\tau}{2}\right) \nonumber\\
		&\quad \Bigg. +\frac{5 l^3}{56\ci^{7/2}}\left(1+\frac{7\alpha\tau}{2}\right)+ \T{\delta} \ci^{3/2} \left( 1 + \frac{3\alpha\tau}{2} \right) \left[ \frac{7l^5}{60\ci^{5/2}}\left(1 + \frac{5\alpha\tau}{2}\right) \right.\nonumber \\
		&\quad \left. + \frac{9l^7}{280\ci^{7/2}}\left(1 + \frac{7\alpha\tau}{2}\right) -\frac{l^3}{3}log \left(\ci\left(1-\alpha\tau\right)\right)\right]\nonumber\\
		&\quad \Bigg. -\frac{l^3}{3}\paren{-\frac{1}{3z_{i}^2}(1+3\sqrt{\T{\delta}}z_{i}\tau)+\frac{\T{\delta}}{2z_{i}}(1+\sqrt{\T{\delta}}z_i \tau)}\Bigg]
	\end{align}
	where $\alpha=\frac{2\sqrt{\T{\delta}}z_{i}^3}{\ci}$. One can clearly see that in the early time of the exotic matter-dominated universe, the leading order of complexity changes as $\tau$. This is the same dependence as the radial position of the brane in the early time of exotic matter dominated universe.\\
	\begin{figure}[ht!]
		\centering
		\includegraphics[width=0.7\linewidth]{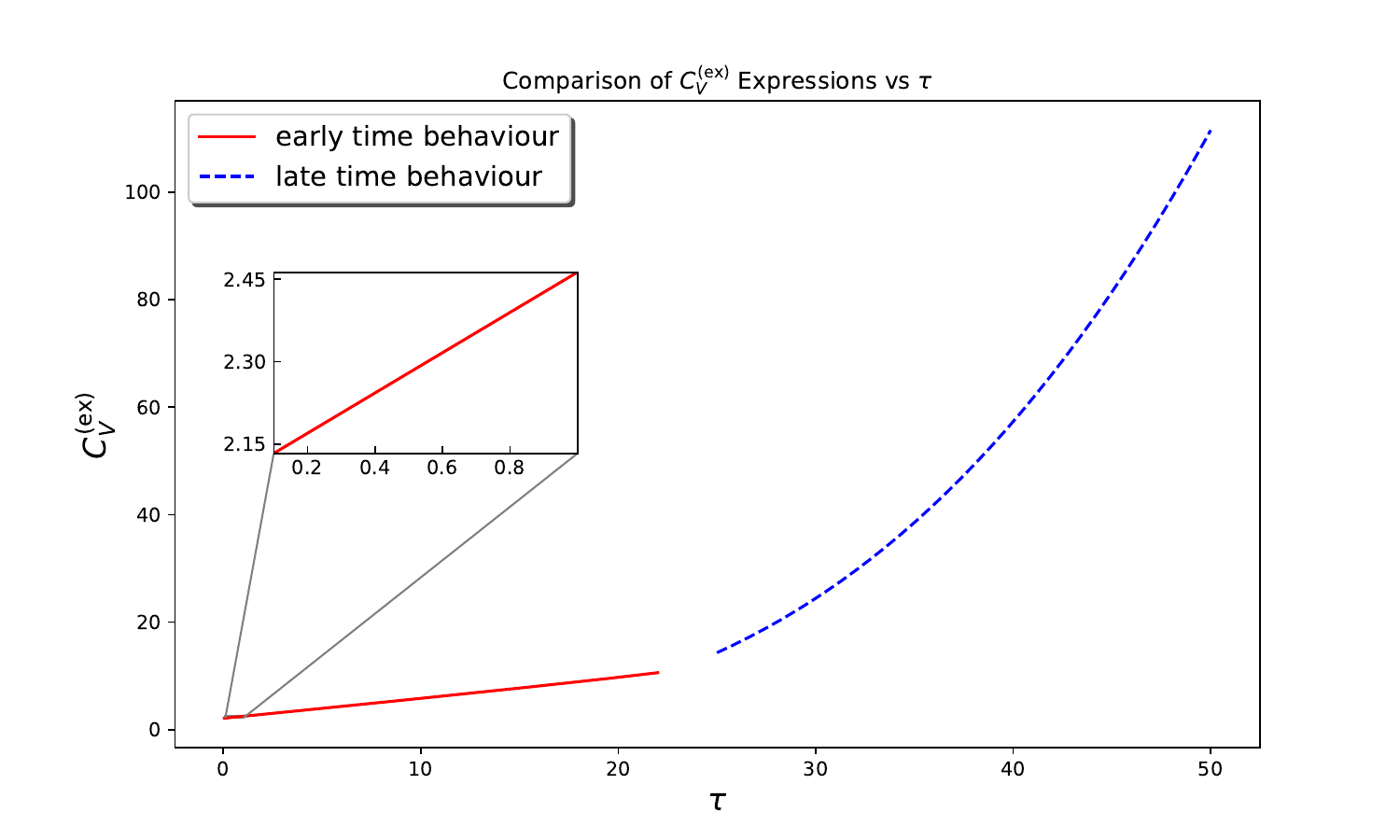}
		\caption{The above plot shows the variation of volume complexity with respect to the cosmological time for the early and late time regimes of the radiation-dominated universe. This plot is done for the parameters $l=2$, $z_i =1$ and $\T{\delta}=0.01$. The dashed plot in red indicates the early-time behavior of holographic complexity of the radiation-dominated era in the early time. The solid line in blue denotes the late-time behavior of the complexity in the radiation-dominated era. Here, the prefactor $\frac{\Omega_2}{4G}$ is set to 1.}
		\label{fig:cvex}
	\end{figure}
	In late time era ($\tau\to\infty$), the brane's position is already given in eq.\eqref{z bar ex late}. Combining the expression of $\z(\tau)$ from eq.\eqref{z bar ex late} in eq.\eqref{z bar ex late}, we get the late time behaviour of complexity in the exotic matter-dominated era, which reads  
	\begin{align}\label{CV_ex_late}
		C^{ex}_{V}&=\frac{\Omega_2}{8\pi G}\Bigg[\frac{1}{3}\paren{1-\frac{3}{2\T{\delta}\tau^2 l^2}}+\frac{1}{10}\paren{1-\frac{5}{2\T{\delta}\tau^2 l^2}}+\frac{5}{56}\paren{1-\frac{7}{2\T{\delta}\tau^2 l^2}}\nonumber\\
		&\quad\Bigg.+\frac{l^3}{3}\paren{\frac{\T{\delta}^{3/2}\tau^3}{3}-\frac{\T{\delta}^{3/2}\tau}{2}}+\T{\delta}l^3 \paren{1+\frac{3}{2\T{\delta}\tau^2 l^2}}\left(\frac{7}{60}\paren{1-\frac{5}{2\T{\delta}\tau^2 l^2}}+\frac{9}{280}\paren{1-\frac{7}{2\T{\delta}\tau^2 l^2}}\right.\nonumber\\
		&\quad\left.-\frac{l^3}{12}log\paren{l\paren{1+\frac{1}{\T{\delta}\tau^2 l^2}}}\right)\Bigg]~.
	\end{align}
	From our result of complexity in the late time era, we can easily see that the leading order of the HSC changes as $\tau^3$. The physical length scale in this case grows as $L\sim \tau$. Therefore, if the volume law holds complexity should scale as $L^3 \sim \tau^3$. The leading order term of the above expression of complexity in late time of exotic matter dominated universe changes as $\tau^3$. Hence, we can safely say that our expression of complexity satisfies the volume law. Comparing this result with the cases of radiation and matter-dominated universe, we have found that in the late time era, the complexity of exotic matter dominated universe grows faster.\\
	Now to understand the dynamics of volume complexity of the exotic matter dominated universe, we have plotted the early and late time behaviour of the volume complexity using eq.(s)\eqref{CV_ex_early},\eqref{CV_ex_late} in Fig.\eqref{fig:cvex}. The plots have been done for $l=2$, $z_i =1$ and $\T{\delta}=0.01$. In a universe dominated by exotic matter, as indicated by the above-mentioned equation, the volume complexity is expected to grow linearly with time, that is, proportionally to $\tau$. The red solid line in the Figure confirms this expectation, which can be verified by examining the zoomed-in inset section. For the late time limit, one can visually confirm from the blue dashed line that for very late times, volume complexity shows $\tau^3$ behaviour. It is also clear that after a certain time, the early-time expression is dominated by the late-time expression of HEE of an exotic matter-dominated universe.
	\vspace{0.8cm}
	\section{Conclusion}\label{sec 7}
	Now we will summarise our findings. In this paper, we have calculated the time-dependent entanglement entropy and complexity of our expanding universe in a holographic manner. We have started the discussion with a brief review of the braneworld model of cosmology.  It is known that after a few microseconds of the Big Bang, a very strongly coupled phase occurred, and it undergoes a confinement and de-confinement phase transition like QCD. This phenomenon encourages us to use the AdS/CFT duality to study the entanglement properties of our universe. As the FLRW model of cosmology deals with non-static spacetime backgrounds, in order to calculate the holographic entanglement entropy one needs to use the HRT formula. The HRT prescription is quite challenging due to the presence of nonlinear coupled differential equations. However, the leading order dependence of the HRT results is given by the RT formula. To explain the properties of the Universe, which is expanding following a power law. The braneworld model of cosmology is used to calculate entanglement entropy and complexity holographically. This model considers that our universe is situated on a brane and different matter sources are realised as the back reaction of different $p$-brane gas geometries in the bulk space. The expanding nature of the universe is realised as the brane's radial motion. The radial motion of the brane with respect to the cosmological time is determined by the second Israel junction condition. For different matter sources (radiation, matter and exotic matter), we have calculated the brane's radial position as a function of cosmological time. We have also given a brief review of this braneworld model. In this braneworld model, we can use the static RT formula and "Complexity=Volume" conjecture to evaluate the expression of holographic entanglement entropy and complexity. Later, to introduce the time dependence, we substitute the brane's position as a function of cosmological time. We have also calculated the early and late time behaviours of all these quantities. In order to do so, we have again used the Israel junction conditions for different matter sources. At first, we have calculated the approximate time-dependent position of the brane for the eternal inflationary era, radiation-dominated, matter-dominated and exotic matter-dominated universe. Then we have put these expressions of the brane's position in the results of HEE and HSC to obtain their early and late time behaviours in different eras. We have shown that for an eternally inflating universe in the absence of any matter source, the HEE changes as $\tau$ and $e^{2H\tau}$. For the radiation-dominated era in early times, HEE changes as $S^{(rad)}_{HEE}\sim\tau^{1/2}$ and in the late time era, HEE changes as $S^{(rad)}_{HEE}\sim\tau$. Similarly, we have also calculated the early and late time dependence of the universe for the matter-dominated era. In early and late times, it changes as $\tau^{2/3}$ and $\tau^{4/3}$ respectively. Also in the presence of exotic matter, the HEE of the universe changes as $\tau$ and $\tau^2$ for early and late times, respectively. We have seen that for radiation, matter and exotic matter-dominated universes, the physical length scales change as $\tau^{1/2}$, $\tau^{2/3}$ and $\tau$ respectively in the late time eras of those respective universes. Hence, in late times, the entanglement entropy of these different matter-dominated eras are proportional to the physical area of the RT surface. In this work, we have derived the forms of holographic entanglement entropy using a perturbative approach for universes dominated by different energy components, such as radiation, matter, and dark energy. Unlike the method used in \cite{Park:2020jio}, our approach follows a different prescription for handling the perturbative expansion and the associated calculations. Despite these prescribed differences, we have shown that our results agree with those of \cite{Park:2020jio} at leading order, which confirms the consistency of our framework. Additionally, we have extended our approach to explore a less commonly studied scenario, namely, a universe dominated by exotic matter, and computed the corresponding holographic entanglement entropy, as presented in Section~\ref{exotic matter}. We have also calculated the volume complexity for an eternally inflating universe and a universe expanding by a power law with radiation, matter and exotic matter-dominated eras. The volume complexity of the eternally inflating universe changes as $\tau$ in early times and $e^{3H\tau}$ in late times. In an eternally inflating universe, the physical length scale changes as $e^{H\tau}$, hence, the late-time behaviour of complexity changes as the physical volume enclosed by the RT surface. We have further investigated the volume complexity across different types of matter-dominated universes, employing the same procedure we used to calculate the holographic entanglement entropy. In case of the radiation-dominated era, the complexity scales as $\tau$ and $\tau^{3/2}$ for early and late time respectively. For the matter-dominated era of the universe, the complexity in the early-time varies as $\tau^{2/3}$ and $\tau^{2}$. Again for a universe with dominant matter as some kind of exotic matter, the complexity changes as $\tau$ and $\tau^3$ in early and late-time scenarios, respectively. These results suggest that in late time era of each of these matter-dominated eras, the complexity is proportional to the physical volume enclosed by the RT surface. We have also provided graphs to clarify the expressions, including a zoomed-in inset to highlight the very early time behavior. From these graphs, it is evident that the growth at later times surpasses that at early times beyond a certain limiting point. This observation supports the consistency of our findings. We would also like to mention that the presence of stiff matter in this model for five dimensions cannot be justified, as it leads to an unphysical p-brane configuration in bulk spacetime. It may happen that stiff matter does not show up in low dimensions as their existence was confirmed in the early universe when the energy scale was very high. This may indicate that we have to do all the calculations in dimensions higher than five to observe the effect of stiff matter in the braneworld model. We leave this topic as a future perspective of our work. Another potential direction for exploration is to investigate the time dependence of various mixed-state entanglement measures, such as mutual information, entanglement wedge cross section, and entanglement negativity, within this braneworld framework. 
	\begin{center}
		\section*{Appendix}\label{appendix}
	\subsection*{Review on four-dimensional FLRW Universe}
    \end{center}
	Here, we will give a short review of the $(3+1)$-dimensional FLRW universe. In the past few decades, several observations\cite{Planck:2015fie,Planck:2018nkj,Planck:2015mrs} have confirmed that the only universe where we live is almost isotropic and homogeneous and the best model to describe is $\Lambda$CDM model \cite{PhysRevD.110.030001,SNLS:2005qlf,WMAP:2006bqn,SupernovaSearchTeam:1998fmf,SupernovaCosmologyProject:2003dcn,SupernovaCosmologyProject:1998vns,SDSS:2003eyi,SupernovaSearchTeam:2004lze}. So it is more accurate to study the $p$-brane gas geometry with an isotropic and homogenous model. 
	The study on the $p$-brane gas geometry requires a introduction of the standard isotropic and homogeneous FLRW universe governed by the metric written in radial coordinates as\cite{Weinberg:1972kfs,Ryden:1970vsj} 
	\begin{equation}
		ds^2=-dt^2+a(t)^2\left[\frac{dr^2}{1-\kappa r^2}+r^2d\Omega_2^2\right]
	\end{equation}
	where $\kappa \in \{0,\pm1\}$ is the spatial curvature index, $a(t)$ is the scaling factor and $\Omega_2$ is the volume of a two sphere. Now in order to derive the Friedmann equations, one can take the following stress-energy tensor of a perfect fluid as 
	\begin{equation}
		T_{\mu\nu}=(\rho+P)u_\mu u_\nu+Pg_{\mu\nu}
	\end{equation}
	where $u_\mu$=(1,0,0,0) in a frame comoving with the fluid. Upon using the above stress-energy tensor and solving Einstein's equations, one gets the Friedmann equations as follows
	\begin{equation}
		\left(\frac{\dot a}{a}\right)^2=H^2=\frac{8\pi G}{3}\rho-\frac{\kappa}{a^2},
	\end{equation}
	\begin{equation}
		\frac{\ddot a}{a}=-4\pi G \left(P+\frac{1}{3}\rho\right).
	\end{equation}
	Another important equation in this regard can be obtained either by combining the above two equations or by the divergence less property of the stress-energy tensor, that is, $\nabla_\mu T^{\mu\nu}=0$. This conditions provides us with the following equation 
	\begin{equation}
		\dot{\rho}+3\frac{\dot a}{a}\left(P+\rho\right)=0~.
	\end{equation}
	Now if we take a linear equation of state $\rho=\omega P$, the above equation takes the form 
	
	\begin{equation}
		\frac{d}{dt}\left[ln (\rho a^{3(1+\omega)})\right]=0~
	\end{equation}
	which implies 
	\begin{equation}
		\rho=\frac{\rho_0}{(a/a_0)^{3(1+\omega)}}
	\end{equation}
	where $\rho_0$ is the energy density when the scale factor is $a_0$.
	Now the first Friedmann equation can be rewritten after defining some new parameters $\tilde a=a/a_0, H^2=\frac{8\pi G }{3}\rho_c,\Omega=\frac{\rho}{\rho_c}$ and using the previous equation 
	\begin{equation}
		\left(\frac{H}{H_0}\right)^2=\frac{\rho}{\rho_{c0}}-\frac{3\kappa}{8\pi G \rho_{c0}}\frac{1}{a^{-2}}=\Omega_0\tilde a^{-3(1+\omega)}-\frac{3\kappa}{8\pi G\tilde a^2a_0^2}
	\end{equation}
	If we define $\frac{-3\kappa}{8\pi Ga_0^2}=\Omega_{k0}$, the form of the  above equation becomes
	\begin{equation}
		\left(\frac{H}{H_0}\right)^2=\Omega_0\tilde a^{-3(1+\omega)}+\Omega_{k0}\tilde a^{-2}~.
	\end{equation} 
	One can now say (at least mathematically) that the spatial curvature is just like an exotic matter which has an equation of state parameter $\omega=-\frac{1}{3}$.\\
	For a spatially flat spacetime, we know that $\Omega_0=1$, this leads us to the following equation
	\begin{equation}
		\frac{\dot{\tilde{a}}}{\tilde{a}}=\frac{H_0}{\tilde a^{\frac{3(1+\omega)}{2}}}~.
	\end{equation}
	Now integrating the above equation, one gets 
	\begin{equation}\label{time dependence}
		\int_0^tdt(\frac{2H_0}{3(1+\omega)})=\int_0^{\tilde a} d\tilde a(\tilde{a}^{\frac{3}{2}(1+\omega)-1})~.
	\end{equation}
	This is now quite clear that $a\sim t^{\frac{2}{3(1+\omega)}}$.
	For radiation dominated universe one has $\omega=\frac{1}{3}$ which means $a\sim t^\frac{1}{2}$. Dust (dark matter) has no pressure means $\omega=0$ gives $a\sim t^\frac{2}{3}$. It is easy to show that for a vacuum universe which has some nonzero spatial curvature or for a flat universe filled with an exotic matter that scales as $\rho\sim a^{-2}$, scale factor scales with time as $a(t) \sim t$. However, this power law behaviour of the scale factor breaks down for $\omega=-1$. In that case, we have to go back to the eq.(\ref{time dependence}) which implies exponential growth of the scale factor, corresponds to the inflationary phase of the universe and also the observed late-time acceleration of the universe.
	\vspace{1cm}
	\section*{Acknowledgments}
	SP would like to thank SNBNCBS for the Junior Research Fellowship.
	GG extends his heartfelt gratitude to CSIR, Govt. of India for funding this research.
\bibliographystyle{JHEP.bst}
\bibliography{ref.bib}	
\end{document}